\documentclass[preprint,aps,amsmath,amssymb,nofootinbib,superscriptaddress,prd,tightenlines,showpacs]{revtex4}
\def\be{\begin{equation}} \def\ee{\end{equation}}
\def\ba{\begin{eqnarray}} \def\ea{\end{eqnarray}}
\def\rme{\textrm{e}}
\usepackage{graphicx}

\def\bra#1{\langle#1\vert}     \def\ket#1{\vert#1\rangle}
\def\ev#1{\langle #1 \rangle}  \def\oo{{\scriptscriptstyle 0}}
\def\dd{\textrm{d}} \def\ii{\textrm{i}} \def\half{{\textstyle{\frac12}}}
\def\kin{_{\textrm{kin}}}  \def\phy{_{\textrm{phy}}} \def\cl{_{\textrm{cl}}}
\def\ha{\hat a}  \def\w{{\omega}}
\def\G{{\Gamma}}  \def\h{\hat} \def\H{{\cal H}} \def\cO{{\cal O}}
\def\g{{\gamma}}  \def\a{{\alpha}} 
\def\U(1){\textrm{U(1)}}  \def\S{{\cal S}}  \def\N{{\cal N}}
\def\z{{\zeta}} \def\u{\underline}  \def\a{\alpha} \def\b{\beta}

\begin{document}
\preprint{\vbox{\baselineskip=12pt \rightline{IGPG-05/04-03}
\rightline{ICN-UNAM-05/02} \rightline{gr-qc/0504052} }}

\title{Semiclassical States for Constrained Systems}

\author{Abhay Ashtekar}
\email{ashtekar@gravity.psu.edu}
\affiliation{Institute for Gravitational Physics and Geometry\\
Physics Department, Penn State, University Park, PA 16802, U.S.A.}
\author{Luca Bombelli}
\email{bombelli@olemiss.edu}
\affiliation{Department of Physics and Astronomy\\
University of Mississippi, University, MS 38677, U.S.A.}
\author{Alejandro Corichi}
\email{corichi@nucleares.unam.mx}
\affiliation{Instituto de Ciencias Nucleares\\
Universidad Nacional Aut\'onoma de M\'exico\\
A. Postal 70-543, M\'exico D.F. 04510, M\'exico}
\affiliation{Department of Physics and Astronomy\\
University of Mississippi, University, MS 38677, U.S.A.}


\begin{abstract}
The notion of semi-classical states is first sharpened by
clarifying  two issues that appear to have been overlooked in the
literature. Systems with linear and quadratic constraints are then
considered and the group averaging procedure is applied to
\emph{kinematical} coherent states to obtain \emph{physical}
semi-classical states. In the specific examples considered, the
technique turns out to be surprisingly efficient, suggesting that
it may well be possible to use kinematical structures to analyze
the semi-classical behavior of physical states of an interesting
class of constrained systems.
\end{abstract}

\maketitle

\section{Introduction}
\label{s1}

\noindent In the standard Hamiltonian descriptions of fundamental
interactions, the canonical variables are subject to constraints.
Notable examples of systems of this type are gauge theories,
general relativity and supergravity. In the gravitational case, a
key question faced by any background-independent approach, such as
loop quantum gravity, is whether specific constructions used to
impose constraints in the quantization procedure  lead to a theory
with  `a sufficient number of semi-classical states'. To analyze
this issue one needs a framework which spells out the precise
meaning of the term `semi-classical states', introduces strategies
to construct them and provides tools to analyze their properties.
The purpose of this paper is to propose such a framework and
illustrate its use with simple examples.

We will restrict ourselves to systems in which the kinematic phase
space $\G$ ---i.e., the initial phase space on which constraints
are ignored--- is a linear space. However, our considerations can
be extended to cases in which $\G$ is a convex subset of a vector
space or an affine space, and these cases cover a very large class
of systems of interest. Our second restriction is more
significant: We will only consider those constraint surfaces
$\bar{\G}\kin$ which are level surfaces of linear or quadratic
functions on $\G$. This assumption will enable us to perform
explicit calculations. The Gauss constraint of the Maxwell theory,
the constraints of linearized gravity, the diffeomorphism
constraint of full geometrodynamics, as well as a number of
constraints often used in the literature for systems with a finite
number of degrees of freedom are of this form. However, our
restriction excludes several important cases, most notably the
Hamiltonian constraint of general relativity.

The basic ideas can be summarized as follows. We will begin with a
set of Dirac observables $\cO_i$ on $\G$ which is sufficiently
large to separate points of the reduced phase space $\h\G$. We
will then choose coherent states $\Psi_\g$ in the kinematical
Hilbert space $\H\kin$ which are peaked at points $\g$ of $\G$ in
the following sense: Expectation values of the quantum observables
$\h\cO_i$ are the values $\cO_i(\g)$ assumed by the classical
observables $\cO_i$ at $\gamma$, with suitably small fluctuations.
The idea is to restrict $\g$ to lie on the constraint surface
$\bar{\G}\kin$ and construct \emph{physical} semi-classical states
$\Psi_\g^{\textrm{phy}}$ by averaging $\Psi_\g$ over the the group
generated by the quantum constraint operator $\h{C}$ on $\H\kin$,
calculate the expectation values and fluctuations of $\h{O}_i$ in
$\Psi_\g^{\textrm{phy}}$, and compare them with those in
$\Psi_\g$.

While the strategy seems natural ---even obvious--- at first, a
priori it is not clear that it would be useful. First, the group
averaging procedure \cite{Marolf} need not result in a state which
has finite and positive norm. Second, even if  it does, and
therefore defines a state $\Psi_\g^{\textrm{phy}}$ in the physical
Hilbert space $\H\phy$, this state may not at all be
semi-classical. For, the original coherent state $\Psi_\g$ in the
kinematic Hilbert space is not infinitely peaked on the constraint
surface $\bar\G$. Indeed, the coherent state $\Psi_\g$ will
necessarily have fluctuations about the constraint surface which,
upon group averaging, could contaminate the expectation values of
the Dirac observables and enhance their fluctuations
uncontrollably. Surprisingly, this does not happen, at least in
simple examples, even when the constraint is quadratic and the
reduced phase space is a genuine manifold so that $\H\phy$ does
not admit any \emph{direct} analogs of the kinematic coherent
states $\Psi_\g$. These examples suggest that there may well be a
general and interesting structure governing the interplay between
semi-classicality and group averaging and a more extensive and
systematic study of the issue may be worthwhile.

The paper is organized as follows. Section \ref{s2} introduces the
notation and briefly recalls a few properties of the standard
coherent states that are repeatedly used in the rest of the paper.
Section \ref{s3} points out certain subtleties associated with the
notion of semi-classicality that are generally ignored in standard
treatments but which are often important in practice, particularly
when dealing with constrained systems. Linear constraints are
discussed in section \ref{s4}. In this case, the expectation
values and fluctuations of a very large class of Dirac observables
in the kinematic coherent states $\Psi_\g$ and their group
averaged versions $\Psi_\g^{\textrm{phy}}$ are identical. An
interesting element is that this result holds even if the
fluctuations of the constraint operator in the state $\Psi_\g$ are
arbitrarily large and indeed even when the point $\g$ does not lie
on the constraint surface $\bar\G$. In section \ref{s5} we
consider quadratic constraints. Now, the analysis can not be made
explicit in the most general situation. Answers involve certain
summations or integrals which are well defined, but
whose properties are not so transparent. To make
explicit comparisons between properties of $\Psi_\g$ and
$\Psi_\g^{\textrm{phy}}$, we consider two specific examples which have
drawn a great deal of attention in connection with quantization of
non-trivially constrained systems and `the issue of time'. In both
cases, the group averaged states $\Psi_\g^{\textrm{phy}}$ are again
semi-classical. In fact, in the more non-trivial case when the
reduced phase space is compact, group averaging actually improves
semi-classicality by \emph{reducing} the fluctuations! These
results are non-trivial in that, in the current analysis, they
arise from delicate cancellations between terms which can not be
foreseen until the very last step. It is these subtleties that
suggest that there may well be a more general underlying structure
yet to be unveiled.

We will conclude this Introduction with two remarks. First, while
there is considerable literature on the coherent state
quantization \cite{jk1,jk2,jk3,jk4,Ashworth}, the focus there is
on constructing a quantum theory of non-trivial systems using
coherent state techniques. Our emphasis is different. We are
primarily interested in \emph{semi-classical} issues in the
resulting quantum theories. Secondly, our goal here is not to
provide a complete and exhaustive analysis of the application of
the group averaging technique to construct candidate
semi-classical physical states. We only wish to present a few
simple examples in the hope that the explicit and rather
intriguing results in these cases may stimulate further research.

\section{Preliminaries}
\label{s2}

\noindent This section is divided into two parts. In the first,
we fix our notation and recall certain facts about standard coherent
states that will be used repeatedly in sections \ref{s4} and
\ref{s5}. In the second, we summarize the group averaging
technique which forms the basis of our construction.

\subsection{Linear phase spaces and coherent states}
\label{s2.1}

\noindent Consider a system with $D$ degrees of freedom, with
linear phase space $\G = \mathbb{R}^{2D}$. It is convenient to use
a canonical basis so that the position vector $\g$ of each phase
space point can be specified by its $2D$ components $(q_i, p_i)$,
with $i = 1,\ 2,\ldots ,\ D$. In examples, the basis will be
adapted to the constraints at hand. The symplectic tensor $\Omega$
on $\G$ is then given by
\be \Omega (\g,\, \tilde{\g}) = p_i \tilde{q}_i - q_i \tilde{p}_i
\, , \ee
where we have used the summation convention for repeated indices.
(Unless stated otherwise, we will follow this convention
throughout this paper.)

In sections \ref{s4} and \ref{s5}, $\G$ will serve as the
kinematical phase space. Since it has a linear structure, the
kinematical quantization can be carried out in a standard fashion.
In the discussion of coherent states, it is convenient to use Fock
quantization. This requires the introduction of a K\"ahler
structure on $\G$, which in turn requires new scales to enable one
to define dimensionless holomorphic coordinates $z_i$. For
definiteness, we will let the new scales $\ell_i$ have dimensions
of length and adapt the initial choice of the canonical basis to
the K\"ahler structure so that $z_i$ is given by
\be z_i := \frac{q_i}{\sqrt2\,\ell_i} + \ii\,
\frac{\ell_i\,p_i}{\sqrt2\,\hbar} \label{zi}\ee
(where there is no summation over $i$ on the right side). The
length scales $\ell_i$ are often provided by the physical
parameters of the system under consideration. For a single
harmonic oscillator, for example, $\ell = \sqrt{\hbar/m\omega}$.
The K\"ahler metric endows the position vectors $\g \equiv
(q_i,p_i)$ in $\G$ with a norm, given by
\be |\g|^2 = z_i\,\bar{z_i}\;. \ee

In quantum theory, $z_i$ are promoted to annihilation operators
$\h{a}_i$ with commutation relations $[\h{a}_i, \h{a}_j^\dag]
=\delta_{ij}\, \h{1}$. Given a specific point $\a$ in $\G$ with
coordinates $(q^\oo_i, p^\oo_i)$, following (\ref{zi}) we will set
\be \a_i := \frac{q^\oo_i}{\sqrt2\,\ell_i} + \ii\,
\frac{\ell_i\,p^\oo_i}{\sqrt2\,\hbar}\, . \label{alphai}\ee
(Again, there is no summation over $i$ on the right side.) Then,
the canonical, normalized coherent state $\Psi_\a$ which is
`peaked' at $\a$ is given by
\ba \ket{\Psi_\a}
&=& \rme^{\a_i \h{a}^\dag_i - \bar\a_i\h{a}_i}\,\ket{0}
\nonumber\\
&=&\rme^{-|\alpha|^2/2} \sum_{n_1,\ldots,n_D=0}^\infty
\frac{(\alpha_1)^{n_1}\cdots(\alpha_D)^{n_D}}{\sqrt{n_1!}
\cdots \sqrt{n_D!}}\, \ket{n_1, n_2, \ldots n_D}\;.
\label{coh} \ea
In the configuration representation, this state can be expressed
as
\be \Psi_\a(q)\, = \prod_{i=1}^D \frac{\rme^{-\ii\,p_i^\oo\,
q_i^\oo/\hbar}} {(\pi\ell_i^2)^{1/4}}\,
\rme^{-(q_i-q_i^\oo)^2/2\ell_i^2}\,\,\,
\rme^{\ii\,p_i^\oo\,q_i/\hbar}\;. \ee
In these states, the expectation values and uncertainties
of the basic canonical variables are given by
\be \bra{\Psi_\a}\,\hat q_i\,\ket{\Psi_\a} = q^\oo_i, \quad
\bra{\Psi_\a}\,\hat p_i\,\ket{\Psi_\a} = p^\oo_i \label{evs}\ee
and
\ba (\Delta \hat q_i)^2 \equiv
\bra{\Psi_\a}\,\hat q_i^2\,\ket{\Psi_\a} -
[\bra{\Psi_\a}\,\hat q_i\,\ket{\Psi_\a}]^2 &=& \half\,\ell_i^2\;,
\nonumber\\
(\Delta \hat p_i)^2 \equiv \bra{\Psi_\a}\,\hat p_i^2\,\ket{\Psi_\a}
- [\bra{\Psi_\a}\,\hat p_i\,\ket{\Psi_\a}]^2 &=&
\half\,\hbar^2/\ell_i^2\;. \label{fluctuations} \ea
Thus, for each value of $i$, the product of uncertainties
$(\Delta\hat q_i)(\Delta\hat p_i)$ is minimized. In fact, the
requirement that (\ref{evs}) and (\ref{fluctuations}) be satisfied
by a state $\Psi_\a$ suffices to restrict $\Psi_\a$ to be a
coherent state, given by (\ref{coh}) modulo a phase factor;
this provides an alternate and more physical definition of a
coherent state. Note also that the length scales $\ell_i$ we
introduced in the specification of the K\"ahler structure have
a direct physical interpretation in terms of coherent states:
$\ell_i/\sqrt{2}$ are the uncertainties in $\hat q_i$ in the
coherent states $\Psi_\a$, canonically defined by the K\"ahler
structure.

Finally, for future use, we note two properties of coherent states
which can be easily verified. First, the scalar product between
two of them is given by
\be \ev{\Psi_\beta|\Psi_\a} = \rme^{-(|\alpha|^2+
|\beta|^2)/2}\,\, \rme^{\bar{\beta}{\a}} \label{sprod}\;. \ee
Second, given any function on the phase space that can be
expressed as a polynomial $F(\bar{z_i}, z_j)$, the corresponding
\emph{normal ordered} operator :$\,F(\h{a}_i^\dag, \h{a}_j)\,$:
has the following expectation value:
\be \ev{\Psi_\beta|:F(\h{a}_i^\dag, \h{a}_j):|\Psi_\a} =
F(\bar{\beta}_i, \alpha_j)\, \ev{\Psi_\beta|\Psi_\a}\;.
\label{expvalue}\ee

\subsection{Group averaging}
\label{s2.2}

\noindent For a large class of physically interesting systems, the
group averaging method provides a technique to extract physical
states starting from kinematical ones. In this sub-section we will
present a pedagogical summary, focusing only on those aspects of
the procedure that are central to sections \ref{s4} and \ref{s5}.
In particular, we will consider a single constraint, although all
our considerations generalize in a straightforward fashion if the
set of quantum constraint operators is Abelian. For a discussion
of more general cases and a treatment of subtleties and technical
caveats see, e.g., Ref.\ \cite{Marolf}.

Let us first suppose that the constraint operator $\h{C}$ on the
kinematical Hilbert space $\H\kin$ is self-adjoint, and that zero
is a discrete point in its spectrum. Then, the kernel of $\h{C}$
is a subspace of $\H\kin$. Therefore, to extract physical states,
one just has to project kinematical states to this subspace. In
the case when the 1-parameter group $\h{U}(\lambda) =
\rme^{-\ii\lambda\h{C}}$ generated by $\h{C}$ on $\H\kin$ provides
a representation of $\U(1)$, the projection procedure can be
explicitly carried out through an integration: Given any $\Psi \in
\H\kin$, set
\be \Psi\phy := \frac{1}{\Lambda}\, \int_0^\Lambda \dd\lambda\,
\rme^{-\ii\lambda\h{C}}\, \Psi\;, \label{proj} \ee
where $\Lambda$ is chosen such that $\rme^{-\ii\Lambda\h{C}} = 1$.
Then $\Psi\phy$ satisfies the constraint, $\h{C}\cdot\Psi\phy =0$,
and is thus a physical state. Since these physical states belong
to $\H\kin$, the scalar product between them is well-defined. In
terms of the general `seed' kinematic states $\Psi$ used in
(\ref{proj}), this can be expressed as
\be \ev{\Psi\phy|\Phi\phy} = \ev{P\,\Psi|P\,\Phi} =
\ev{P\,\Psi|\Phi} = \frac{1}{\Lambda} \int_0^\Lambda \dd\lambda\,
\ev{\rme^{-\ii\lambda\h{C}}\, \Psi|\Phi}\;. \label{ip1}\ee
Thus we have reformulated the expressions of the projection
operator and the inner product using averages over the group
generated by the constraint. This reformulation can be readily
carried over to the non-trivial case where the kernel of the
constraint operator does \emph{not} belong to $\H\kin$.

Let us now consider the more difficult case when $\h{U}(\lambda)=
\rme^{-\ii\lambda\h{C}}$ provides a representation of the group
$\mathbb{R}$ on $\H\kin$. In this case, one starts by choosing a
suitable dense sub-space $\S$ of $\H\kin$ (specified below). Then,
for each $\Psi\in \S$, one defines a `generalized bra' via
\be (\Psi\phy| := \frac{1}{K}\, \int \dd\lambda\,
\bra{{\rme^{-\ii\lambda\h{C}}\, \Psi}} \;, \label{extractor}\ee
where the integral is now over the entire real line and $K$ is a
constant (independent of $\Psi$) whose value can be chosen
conveniently. While for any fixed $\lambda$,
$\bra{{\rme^{-i\lambda\h{C}}\, \Psi}}$ is a well-defined bra, the
integral fails to have a finite norm in $\H\kin$. \emph{The round
bracket in $(\Psi\phy|$ emphasizes the fact that now the physical
states do not belong to $\H\kin$.} Rather, they represent
`distributions' in the following sense: the sub-space $\S$ of
$\H\kin$ is chosen such that $(\Psi\phy|$ is an element of the
(topological) dual $\S^\star$ of $\S$ (with respect to a topology
which is finer than that of $\H\kin$). The meaning of
(\ref{extractor}) is simply that the action of this element of
$\S^\star$ on any $\Phi$ of $\S$ is given by
\be (\Psi\phy\ket{\Phi} = \frac{1}{K}\, \int \dd\lambda\,
\ev{\rme^{-\ii\lambda\h{C}}\, \Psi|\Phi}\;.
\label{distribution}\ee

Since $\Psi\phy$ no longer lies in $\H\kin$, one now needs a new
prescription for defining the scalar product between two physical
states. The idea is to use expression (\ref{ip1}) as a motivation
and set the physical inner product to be
\be (\Psi\phy|\Phi\phy) = (\Psi\phy\ket{\Phi}\;, \label{ip2}\ee
so that the norm of physical states is given by%
\footnote{This procedure can be heuristically understood as
follows. (\ref{extractor}) extracts from $\Psi$ a physical state
$\Psi_{\textrm{phy}} \in \S^\star$. This extractor $\hat{E}$ can be
formally thought of as $\hat{E} \Psi = \delta(\hat C) \Psi$.
Therefore, the naive definition $\ev{\Psi\phy|\Psi\phy} =
\ev{\delta(\h{C})\Psi| \delta(\h{C}) \Psi}$ of the norm that one
may first think of is divergent. In the correct definition,
(\ref{norm}), one of the two delta-distributions is simply
dropped, thereby removing the obvious infinity.}
\be \Vert\Psi\phy\Vert^2 = \frac{1}{K} \int \dd\lambda\,
\ev{\rme^{-\ii\lambda\h{C}}\, \Psi|\Psi}\;. \label{norm}\ee
Thus, in the present case when the group generated by $\hat{C}$ of
$\H\kin$ is non-compact, the group averaging procedure provides
the inner product in the physical Hilbert space only up to an
overall multiplicative constant $K$. Its value is generally chosen
to remove an overall irrelevant constant from the expression of
the physical inner product.

For a general choice of the initial subspace $\S$, there is no
guarantee that the norm (\ref{norm}) would be finite and positive.
The `art' in the group averaging procedure lies in selecting a
dense subspace $\S$ of $\H\kin$ such that: i) the right side of
(\ref{distribution}) is well-defined for all $\Psi, \Phi \in \S$;
i.e., $\Psi\phy$ is a well-defined distribution over $\S$; and,
ii) the norm (\ref{norm}) of each $\Psi\phy$ is non-negative,
vanishing if and only if $\Psi\phy$ vanishes. The procedure
succeeds in its goal of constructing the physical Hilbert space
only if such a $\S$ can be located. If more than one viable
candidate exist, the resulting quantum theories could well be
inequivalent. However, for the semi-classical issues discussed in
this paper the situation is simpler, in that there is a natural
choice which can be shown to be viable: we will let $\S$ be
spanned by finite linear combinations of kinematical coherent
states.

For later use, we introduce the following notation: the physical
expectation value of an observable $\hat\cO$ will be denoted by
$\ev{\hat\cO} \phy =
({\Psi}\phy\ket{\hat\cO\Psi}/({\Psi}\phy\ket{\Psi}$. In the
explicit examples, we will encounter both types of constraints: in
some cases $\h{U}(\lambda) = \rme^{-\ii\lambda\h{C}}$ will provide
a representation of $\U(1)$ and in others of $\mathbb{R}$.

\emph{Remark:} Klauder \cite{jk1} has introduced an alternative
procedure to quantize constrained systems. When the group
generated by constraint operators on $\H\kin$ is non-compact, the
idea is to extract physical states in two steps. In the first, one
projects on the subspace of $\H_{\rm kin}$ corresponding to a
portion $[-\delta, \delta]$ of the spectrum of the constraint
operator. In the second, one takes the limit as $\delta$ goes to
zero. A case-by-case study is needed to check if the second step
can be carried out successfully. This is analogous to the fact, in
the group averaging procedure, that there is no a priori guarantee
that the candidate physical inner product would be positive
definite. However, to our knowledge, the precise relation between
the two procedures has not been studied.

\section{Semi-classicality}
\label{s3}

\noindent Even for simple unconstrained systems, the textbook
treatments of semi-classicality generally overlook two points
which are important to our analysis. We will first discuss these,
and then consider issues relevant to constrained systems.

Fix a point $\a$ in a linear phase space $\G$ with coordinates
$(q^\oo_i,p^\oo_i)$. Our task is to spell out what we mean by a
semi-classical quantum state which is `peaked at this classical
state'. The intended meaning is intuitively clear and, although it
is not always stated explicitly, one generally has the following
idea in mind: a semi-classical state $\Phi_{\a}$ should be such
that, for all well-behaved functions $F(q_i,p_i)$ on phase space,
the expectation values $\ev{\Phi_{\a}|\h{F}|\Phi_{\a}}$ are close
to $F(q^\oo_i, p^\oo_i)$ and the fluctuations small. However,
such semi-classical states simply don't exist unless the class
of observables is greatly restricted.

Let us discuss this issue in some detail. Take a harmonic
oscillator. Then, one generally takes coherent states $\Psi_\a$ of
(\ref{coh}) as representing semi-classical states peaked at the
point $\alpha$ of the phase space $\G$. Indeed, if one restricts
oneself to the set, say, of three key observables, $q,p,H$ of the
system, where $H$ is the Hamiltonian, $\Psi_\a$ would satisfy the
above criteria (if the words `close to' and `small' are
interpreted appropriately; see below). However, if the set also
includes the observable $\rme^{H/\epsilon}$ with $\epsilon \le
\hbar\omega$, coherent states would strongly violate the criteria.
Of course, for the harmonic oscillator the new observable is
rather strange and it is difficult to justify its inclusion in the
list on physical grounds.

However, there is in fact an interesting situation in which the
analogous observable is of direct physical interest. This is
provided by the quantum theory of the Einstein-Rosen waves in
4-dimensional general relativity \cite{kuchar}. This system is
equivalent to the axi-symmetric sector of 2+1 dimensional general
relativity coupled to a scalar field \cite{ap}. In the Hamiltonian
framework, one can arrange matters so that the true degrees of
freedom are coded in the scalar field $\phi$ propagating on a
fiducial 3-dimensional Minkowski space. The physical scalar field
can be identified with $\phi$ and the physical metric coefficients
are also completely determined by it. Outside the support of
$\phi$, the only non-trivial metric component $g$ is given by $g =
\rme^{GH}$, where $H$ is the Hamiltonian of the scalar field in
Minkowski space and $G$ is Newton's constant in 2+1 dimensions.
Since the true degree of freedom is in $\phi$, one can carry out a
Fock quantization and represent all interesting observables,
including $\h{g}$, as operators on this Fock space. Let us now
consider the problem of defining semi-classical states for this
system. If the list of observables of interest includes only the
(smeared) scalar field and its (free-field) Hamiltonian, the
standard coherent states are good semi-classical states. However,
if the list includes also the metric component $\hat{g}$, they are
not \cite{aa}! Conversely, one may construct states which are
semi-classical for $\hat{g}$ but these then fail to be
semi-classical for the (smeared) field $\hat\phi$ \cite{gp}. Thus,
there is no canonical notion of semi-classicality for the system,
independent of one's choice of observables.

\emph{The first lesson then is that to ask for semi-classical
states, one must first specify a class of observables for
which the states are to be semi-classical.} A state may be
semi-classical for one choice but not for another.\footnote{
Heuristically, this can be understood in the following terms.
The phase space structure provides a natural symplectic measure
and, in the Bargmann-type representation, semi-classical states
peaked at a point $\gamma$ are concentrated in a neighborhood
of size $\hbar^D$. However, the `shape' of this neighborhood
can vary. For one shape, they would be good semi-classical
states for one set of observables while for another shape,
they would be semi-classical for another set of observables.
We thank Carlo Rovelli for this remark.}

The second subtlety has to do with the notion of fluctuations. The
requirement that the fluctuations of an observable $\h{F}$ in a
state $\Psi$ be small is generally formulated as
\be \frac{(\Delta\h{F})_{\Psi}^2} {|\ev{\Psi|\h{F}|\Psi}|^2}
\,\,\equiv \,\,\frac{\bra{\Psi}\hat F^2\ket{\Psi} -
[\bra{\Psi}\hat F\ket{\Psi}]^2}{|\ev{\Psi|\h{F} |\Psi}|^2} <
\delta^2 \;, \label{wrong} \ee
where $\delta$ denotes the `tolerance' one wishes to allow. There
is, however, the following problem with this proposal: if the
expectation value $\ev{\h{F}}_\Psi$ vanishes, the requirement can
never be met.%
\footnote{One might think that this problem can be trivially
overcome simply by replacing $\h{F}$ with $\h{F} + c \h{1}$ for
a suitable constant $c$ (of appropriate dimension). This strategy
has two problems. First, by choosing the constant to be
sufficiently large, any state can be made to satisfy
(\ref{wrong}). More importantly, the example below shows that one
continues to run into a problem no matter what constant one
adds.}
For constrained systems this issue is especially serious because
we would, in particular, like to consider kinematical coherent
states which are peaked at a point on the constraint surface with
only a small spread and compare them with physical semi-classical
states. Criterion (\ref{wrong}) would forbid us from considering
such states.

The following simple example shows that, for a number of
semi-classical considerations, (\ref{wrong}) is indeed an
incorrect way to encode the idea that the fluctuations of
$\hat F$ are small. Consider a macroscopic, 1-dimensional
harmonic oscillator, such as a pendulum, with position $\h{q}$ as
one of the observables used to test semi-classicality of states.
Consider a point $(q^\oo,p^\oo)$ on the phase space, with real and
imaginary parts of $\a$ (of Eq (\ref{alphai})) large compared to
1. Then the coherent state $\Psi_\a$ of (\ref{coh}) is a good
semi-classical state by any reasonable criterion (and also
satisfies (\ref{wrong}). Now, as is well known, under time
evolution the coherent state remains coherent and its peak simply
follows the classical trajectory of the oscillator. Furthermore,
the uncertainty $\Delta \h{q}$ is time independent. Yet, since the
classical trajectory passes through phase space points at which
$q$ vanishes periodically, criterion (\ref{wrong}) would have us
say that the state violates semi-classicality at those times.
Clearly, this is just wrong! Physically, since the state is
semi-classical initially and does not spread, it is semi-classical
at all times. Thus, we need to modify the criterion (\ref{wrong}).

Let us use direct physical considerations to develop the
appropriate replacement. Given any state, if one is allowed to
make arbitrarily accurate measurements of any one observable, one
would invariably find deviations from the classical behavior.
Thus, a quantum state can be well approximated by a classical one
only if the experimental accuracy is limited. To test
semi-classicality, we must supply information about these
experimental limitations, i.e., tolerances which are fixed at the
outset. We will need two sets of numbers, one specifying the
tolerance in the accuracy of the expectation value, and the other
one that in fluctuations.

These considerations lead us to a specific notion of
semi-classicality that will be used in sections \ref{s4} and
\ref{s5}. \emph{A state $\Psi_\a$ will be said to be peaked at the
point $\a \in \G$ and semi-classical with respect to a given set
of observables ${F}_i$ if}
\be |\ev{\Psi_\a|\h{F}_i|\Psi_\a} - F_i(\a)| < \epsilon_i \quad
\textrm{and} \quad (\Delta\h F_i)_{\Psi_\a} < \delta_i\;,
\label{correct1}\ee
\emph{where $\epsilon_i$ and $\delta_i$ are pre-specified
tolerances determined by the desired experimental accuracy.}

Finally, let us consider constraints. As explained in section
\ref{s1}, the idea is to use the group averaging technique to
extract physical semi-classical states $\Psi^{\textrm{phy}}_\a$
---i.e., semi-classical states which are annihilated by the
constraint operator--- starting from standard coherent states
$\Psi_\a$ in $\H\kin$. Since the notion of semi-classicality is
relative to a set of observables, we will begin by fixing the set
of Dirac observables $\cO_i$ of interest, together with tolerances
$\epsilon_i$ and $\delta_i$. By definition, the physical states
$\Psi^{\textrm{phy}}_\a$ will be semi-classical if they satisfy
(\ref{correct1}). The issue then is: Can we make a suitable choice
of $\Psi_\a$ that will guarantee that the $\Psi^{\textrm{phy}}_\a$
are semi-classical? An example of a sufficient condition for the
answer to be affirmative is
\be |\ev{\h{\cO}_i}\phy - \ev{\Psi_\a|\h{\cO}_i|\Psi_\a}| <
\half\,\epsilon_i \quad \textrm{and} \quad |(\Delta\h
\cO_i)_{\Psi^{\textrm{phy}}_a} - (\Delta\h \cO_i)_{\Psi_a}| <
\half\,\delta_i\; . \label{correct2}\ee
For, if this were the case, we would just need to select the
kinematic coherent states $\Psi_\a$ to satisfy our criterion
(\ref{correct1}) for $\cO_i$, with tolerances $\epsilon_i/2$ and
$\delta_i/2$ (assuming the $\delta_i/2$ are compatible with the
uncertainty relations). In the next two sections we will see that
these conditions are met, even in cases where there is no obvious
a priori reason for this to happen.

\section{Linear constraints}
\label{s4}

\noindent Let us now consider constraints of the type
\be C := K_i q_i + \tilde{K}_i p_i - \Delta = 0\;, \label{lin}\ee
where $K_i, \tilde{K_i}$ and $\Delta$ are any real constants. For
simplicity but without loss of generality we will assume that $C$
is dimensionless. We will first consider a single constraint
(\ref{lin}) and show that group averaging of the kinematical
coherent states $\Psi_\a$ provide physical semi-classical states.
This discussion can be easily extended to incorporate a set of
\emph{commuting} linear constraints. In the second part of this
section, we will illustrate these constructions by applying them
to the Gauss constraint of the quantum Maxwell field in Minkowski
space.%
\footnote{However, since our primary interest lies in the relation
between kinematical and dynamical semi-classical states, we will
refrain from digressing into infinite dimensional subtleties.
Readers who are familiar with Gaussian measures on infinite
dimensional spaces should be able to fill in the details easily.
Those who are not so familiar, can follow the reasoning using the
close analogy with the finite dimensional example discussed in
section \ref{s4.1}.}
Constraints of linearized gravity can be treated in a completely
analogous fashion.

\subsection{General structure}
\label{s4.1}

\noindent Using the definition (\ref{zi}) of the complex
coordinate $z_i$, the constraint (\ref{lin}) can be written as
$C:= \bar{\kappa}_i z_i + \kappa_i \bar{z}_i - \Delta =0$, where
the complex numbers $\kappa_i$ are related to $K_i$, $\tilde{K}_i$
in the obvious manner. Given any coherent state $\Psi_\a$ in the
kinematical Hilbert space, where $\a$ is not necessarily on the
constraint surface, it is easy to verify that
\be \h{U}(\lambda) \Psi_\a := \rme^{-\ii\lambda\h{C}}\, \Psi_\a =
\rme^{-\ii\lambda\,C(\alpha)}\,\Psi_{\a{(\lambda)}}\;,\ee
where $C(\a)$ is the value of the classical constraint $C$ at $\a$
and $\a_j(\lambda) = \a_j - \ii\lambda\kappa_j$ with $\a_j$
labelling the initial phase space point $\a$ as in (\ref{alphai}).

Thus, apart from a phase factor, the image of the coherent state
$\Psi_a$ is another coherent state $\Psi_{\a(\lambda)}$ whose peak
is displaced by $-\ii\lambda\kappa_j$. The physical state
$\Psi_\a^{\textrm{phy}}$ is just a (continuous) superposition
of these displaced coherent states.

To make the properties of $\Psi_\a^{\textrm{phy}}$ and the expressions
of Dirac observables transparent, it is convenient to tailor the
initial choice of canonical coordinates $(q_i,p_i)$ to the given
constraint. It is obvious that we can always orient our basis so
that we have $K_iq_i + \tilde{K}_ip_i = q_1$, and the constraint
reduces then to the simple form
\be C = q_1 - \Delta =0\;. \label{q1}\ee
For simplicity, let us choose all $\ell_i$ to be equal. Then in
the $q$-representation, the action of $\h{U}(\lambda)$ on
$\Psi_\a$ further simplifies
\be \h{U}(\lambda) \Psi_\a(q) = \N_\a\,\,
\rme^{\ii\lambda\Delta}\,\, \rme^{\ii(-\lambda
q_1+p^\oo\cdot\,q/\hbar)}\,\, \rme^{-|q-q^\oo|^2/2\ell^2}\;, \ee
where $\N_\a = \rme^{-\ii p^\oo\cdot\,q^\oo/\hbar}/
(\pi\ell^2)^{D/4}$  is the normalization constant of the initial
coherent state, and $|q-q^\oo|^2 = (q_i-q_i^\oo)(q_i-q_i^\oo)$.
Thus, now the only shift in the peak of the coherent state is in
the first component of the momentum: $p^\oo_1 \rightarrow p^\oo_1
- \lambda$. Consequently, (apart from an overall constant phase
factor) the operator $\h{U}(\lambda)$ simply moves the peak of the
coherent state along the gauge orbit generated by the classical
constraint in the phase space $\G$. This direct interplay between
classical and quantum theories is tied to the fact that the
constraint is so simple. The physical state can also be readily
calculated. We fix the overall rescaling freedom in
(\ref{extractor}) by setting $K = 2\pi/(\pi\ell^2)^{1/2}$
for later convenience, and we obtain
\ba \Psi_\a^{\textrm{phy}}(q) &:=&
\frac{(\pi\ell^2)^{1/2}}{2\pi}\, \int \dd\lambda
\,\h{U}(\lambda) \Psi_\a(q)\,\nonumber\\
 &=& (\pi\ell^2)^{1/2}\,\N_\a\, \delta(q_1-\Delta)\,
 \rme^{\ii(p^\oo\cdot\,q)/\hbar}
\,\rme^{-|q-q^\oo|^2/2\ell^2}\nonumber \\
&=& (\pi\ell^2)^{1/2}\, \delta(q_1-\Delta) \Psi_\a(q)
 \;. \label{phy} \ea
Thus, just as one would have expected from section \ref{s2.2},
$\Psi_\a^{\textrm{phy}}(q)$ is a genuine distribution; it
\emph{fails} to belong to the kinematical Hilbert space $\H\kin$.
Nonetheless, because the constraint $\h{C}$ is diagonal in the
$q$-representation, the group averaging procedure has a simple
interpretation: modulo a state-independent multiplicative
constant, the physical state is just the restriction of the
initial coherent state to the constraint surface $q_1 = \Delta$.

To treat more complicated situations such as the Maxwell theory
discussed in section \ref{s4.2}, it is useful to spell out the
interplay between the reduced phase space quantization and the
group averaging method used here. Since the constraint $q_1-\Delta
=0$ is so simple, there is a natural projection from the phase
space $\G$ and the the $2(D-1)$ dimensional reduced phase space
$\u\G$, spanned by $\u{q}_I, \u{p}_I$, where $I =2,\ldots D$.
Therefore, with every kinematical coherent state $\Psi_\a$, one
can naturally associate a coherent state $\u\Psi_{\u\a}$ in the
reduced phase space quantization. Finite linear combinations of
$\Psi_\a^{\textrm{phy}}$, where $\a$ lies on the constraint
surface, span a dense sub-space of $\H_{\textrm{phy}}$. For these
physical states, it is straightforward to check that the inner
products $\ev{\Psi_\a^{\textrm{phy}}|\Psi_\b^{\textrm{phy}}}$ in
the group averaging method are equal to the inner products
$\ev{{\u\Psi}_{\u\a}|{\u\Psi}_{\u\b}}$ in the reduced phase space
quantization:
\be \ev{\Psi_\a^{\textrm{phy}}|\Psi_\b^{\textrm{phy}}} :=
(\Psi_\a^{\textrm{phy}}\ket{\Psi_\b}=
\ev{{\u\Psi}_{\u\a}|{\u\Psi}_{\u\b}}\;. \label{relation1} \ee

Furthermore, in this example, the physical inner product is also
simply related to the kinematical inner product in the following
sense. Suppose as before that $\a,\b$ lie on the constraint
surface. Let $\hat{\a}, \hat\b$ be points on a `gauge fixed
surface' (so that the corresponding configuration and momentum
coordinates satisfy $\hat{q}^\oo_1 = \Delta$ and $\hat{p}^\oo_1 =
\Delta'$ for some fixed $\Delta'$), which are related to $\a,\b$
via $\a_I = \hat{\a}_I, \b_I =\hat{\b}_I$ for $I=2,\ldots D$.
Then, we also have:
\be  \ev{\Psi_\a^{\textrm{phy}}|\Psi_\b^{\textrm{phy}}} =
\ev{\Psi_{\hat\a}| \Psi_{\hat\b}}\; , \label{relation2} \ee
where the inner product on the right side is that in the
kinematical Hilbert space. Note, however, that \emph{this equality
of inner products does not imply that $\Psi_{\hat\a}$ are physical
states;} they do not satisfy the constraints. As emphasized above,
because of their distributional character (see (\ref{phy})), none
of the physical states belong to the kinematical Hilbert space
while $\Psi_{\hat{\a}}$ clearly do. Nonetheless, (\ref{relation2})
provides a convenient tool to evaluate scalar products between the
physical semi-classical states. In particular, it implies that the
physical states under consideration are normalized:
\be \Vert\Psi\phy\Vert^2 =
\int\dd^Dq\;\bar{\Psi}_\a^{\textrm{phy}}(q)\Psi_\a(q) = 1 \;. \ee
This is a consequence of our choice $K = 2\pi/(\pi
\ell^2)^{\frac{1}{2}}$ of the overall constant $K$ in
(\ref{extractor}).

Let us now turn to observables. The form (\ref{q1}) of the
constraint immediately provides us with a complete set of Dirac
observables which strongly commute with $C$: $q_I, p_I$, where $I
= 2$, $3,\ ...,\ D$. We will work with a larger set, consisting of
general polynomials $F(\bar{z}_I, \bar{z}_J)$ and their
\emph{normal ordered} quantum versions $\h{F} =\; :F(\a_I^\dag,
a_J):$. As noted in section \ref{s2}, the expectation values of
these operators in the kinematic coherent states $\Psi_\a$ are
just the values $F(\bar{\a}_I, \a_J)$ of the classical functions
$F$, evaluated at the points $\a$ of the phase space:
\be \ev{\Psi_\a| \h{F}\,|\Psi_\a} = F(\bar{\a}_I, {\a}_J)\;. \ee
To calculate fluctuations, by moving all annihilation operators in
the expression of $\h{F}$ to the right of all creation operators
and keeping track of the commutator terms that result, one can
express $\h{F}$ as a linear combination $\hat G$ of products of
normal ordered creation and annihilation operators. Hence, the
fluctuations are given by
\be (\Delta{\h{F}})^2_{\a} = \ev{\h{F}^2}_{\a} - (\ev{\h{F}}_\a)^2
= G(\bar{\a}_I, {\a}_J) - (F(\bar{\a}_I, {\a}_J))^2\,. \ee
Since $\h{F}$ and $\h{G}$ do not involve $\h{a}_1$ and
$\ha_1^\dag$, it is easy to calculate the expectation values and
fluctuations also in the (normalized) physical states
$\Psi_\a^{\textrm{phy}}(x)$. One obtains
\be {\ev{\h{F}}_\a^{\textrm{phy}}\, = \, F(\a_I, \bar{\a}_I), \quad
\textrm{and} \quad  (\Delta{\h{F}}_{\a}^{\textrm{phy}})^2} = G(\a_I,
\bar{\a}_I) - (F(\a_I, \bar{\a}_I))^2\,. \ee
Thus, the expectation values and fluctuations of our large class
of Dirac observables are identical in the kinematical coherent
states and in the physical states obtained from them by group
averaging. Therefore, if we ensure that the $\Psi_\a$ are
semi-classical for the given set of Dirac observables, the
$\Psi_\a^{\textrm{phy}}$ will also satisfy our semi-classicality
criteria.

Given the form of $\Psi_\a^{\textrm{phy}}$ and the form of Dirac
observables, this overall result could have been anticipated.
Nonetheless, there \emph{are} two aspects which are rather
surprising, at least at the outset. The results hold even when i)
the initial choice of $\Psi_\a$ is such that the fluctuation in
the constraint is arbitrarily large; and, ii) the point $\a$ does
not even lie on the constraint surface.

\subsection{Example: Gauss constraint for the Maxwell field}
\label{s4.2}

\noindent To make the general construction of section \ref{s4.1}
concrete we will briefly discuss a field theory example in which
there is an infinite number of commuting, linear constraints.
Also, to provide a complementary perspective, we will use the Fock
rather than the configuration representation.

Let $M$ denote a $t$ = constant plane in Minkowski space-time. The
kinematical phase space $\G$ of the Maxwell theory consists of
pairs $(A_a(x), E^a(x))$ of suitably regular fields on $M$. The
constraint surface is defined by $D_aE^a =0$. In the Fock
quantization, it is more convenient to work with the Fourier
components of these fields. Let us introduce an orthonormal basis
$(\h{k}^a, m^a, \bar{m}^a)$, where $\h{k}^a$ is the unit radial
vector and $m^a, \bar{m}^a$ provide a normalized `spin-dyad' on
each 2-sphere to which $\h{k}^a$ is normal. Then, the phase space
can be coordinatized by pairs $(q_i(k), p_i(k))$ of fields in
momentum space, given by
\ba A_a(x) &=& \frac{1}{(2\pi)^{3/2}}\,\, \int \dd^3k\, \rme^{\ii
k\cdot x}\, \left(q_1(k)\,\h{k}_a + q_2(k)\,m_a+ q_3(k)\,
\bar{m}_a\right)\nonumber\\
E^a(x) &=& - \frac{1}{(2\pi)^{3/2}}\,\, \int \dd^3k\, \rme^{\ii
k\cdot x}\, \left(p_1(k)\,\h{k}^a + p_2(k)\,m^a + p_3(k)\,
\bar{m}^a\right). \ea
The Poisson brackets $\{E^a(x), \, A_b(y)\} =
\delta^a_b\,\delta^3(x,y)$ then imply that the fields $q_i(k)$,
$p_j(k)$ are canonically conjugate in the sense that
\be \{q_i(-k), p_j (k')\} = \delta_{ij}\,\delta^3(k,k')\;.\ee

The standard K\"ahler structure is given by the positive and
negative frequency decomposition. The holomorphic coordinates
$z_i(k)$ are now given by
\be z_j(k) = \frac{1}{\sqrt{2}}\,\bigg( \sqrt{|k|}\, q_j(k) -
\frac{\ii} {\hbar \sqrt{|k|}}\, p_j(k)\bigg)\, .\ee
The Gauss law $D_aE^a= 0$ is equivalent to $p_1(k) =0$ which, in
turn, can be recast as an infinite set of commuting constraints,
\be C_f(k) := \int {\dd^3k}\, \bar{f}(k)(z_1(k) - \bar{z}_1(k)) =0
\;, \ee
one for each regular function $f(k)$ in the momentum space (e.g.,
an element of the Schwartz space in $\mathbb{R}^3$). Together,
these constraints are equivalent to the requirement that $E^a$
have no longitudinal modes $p_1(k)$. A complete set of Dirac
observables is therefore given by arbitrary real-valued
functions of the transverse modes $z_I(k), \bar{z}_I(k)$,
where $I=2,\,3$.

The creation and annihilation operators correspond to the
classical functions $z_i(k)$ and $\bar{z}_j(k)$ respectively. They
satisfy the usual commutation relations:
\be [\ha_i(k),\, \ha^\dag_j(k')] = \delta_{ij}\,\delta^3(k,k')\,
\h{1}\;.\ee
The kinematic Hilbert space $\H\kin$ is the Fock space obtained
by operating repeatedly with the creation operators on the vacuum
state $\ket{0}$. For each point $\a$ in the phase space, a
coherent state $\Psi_\a$, peaked at $z_i(k) = \a_i(k)$, can now
be constructed in $\H\kin$ following the procedure outlined in
section \ref{s2.1}:
\be \ket{\Psi_\a} = \rme^{\int (\dd^3k/|k|)\, ({\a}(k)\cdot
\h{a}^\dag(k) - \bar{\a}(k)\cdot \h{a}(k))}\, \ket{0}\;.\ee

For each mode $k$, we now have a linear constraint $p_1(k)=0$.
Hence, the passage to quantum theory is structurally similar to
that in section \ref{s4.1}. Let us therefore begin with a
kinematic coherent state $\Psi_\a$ and apply the group averaging
procedure to it. The physical state $(\Psi_a^{\textrm{phy}}|$ is a
well-defined distribution over the sub-space $\S$ of the Fock
space $\H\kin$ spanned by finite linear combinations of coherent
states. It does not belong to $\H\kin$ because its kinematical
norm diverges. However it is straightforward to calculate the
action of the distribution $(\Psi_\a^{\textrm{phy}}|$ on elements
$\ket{\Psi_\b}$. In the interesting case when $\a$ and $\b$ lie on
the constraint surface, the action is simple to write down (see
(\ref{relation2})):
\be (\Psi_\a^{\textrm{phy}}\ket{\Psi_\beta} =
\ev{\Psi_{\h\a}|\Psi_{\h{\beta}}}\;, \label{relation3}\ee
where $\h\a_1(k) =0,\,\, \h\a_I(k) = \a_I(k)$ and $\h\beta_1(k)
=0,\,\, \h\beta_I(k) = \beta_I(k)$.
This equality is useful in computations: $\Psi_\a^{\textrm{phy}}$,
with $\a$ lying on $\bar\G$ span a dense subspace of $\H_{\textrm{
phy}}$ and the physical inner product between these states is
given by:
\be  \ev{\Psi_\a^{\textrm{phy}}|\Psi_\b^{\textrm{phy}}} =
\ev{\Psi_{\hat\a}| \Psi_{\hat\b}}\; . \label{relation4} \ee
However, as emphasized in section \ref{s4.2} after
(\ref{relation2}), it does \emph{not} imply that physical states
form a sub-space of $\H\kin$. While states $\Psi_{\h\a}$ in
$\H\kin$ are peaked at points on the constraint surface of the
phase space, they have fluctuations also away from the surface.
They do \emph{not} satisfy the quantum constraint.

Since Dirac observables $\cO$ in our complete set depend only on
$z_I$, it follows from the arguments given in section \ref{s4.1}
that the expectation values and fluctuations of $\h\cO$ in
$\Psi_\a$ and $\Psi_\a^{\textrm{phy}}$ are identical.

\section{Quadratic constraints}
\label{s5}

\noindent We will now consider constraints for which $\bar\G$ is
the level surface of a quadratic function on $\G$. As in section
\ref{s3}, without loss of generality, we will suppose that $C$ is
dimensionless. Furthermore, for technical simplicity we will make
a further restriction: the constraint function will be assumed to
be of the form:
\be C(q_i,p_i) := S_{ij}\, q_iq_j + \Lambda\, S_{ij}\, p_ip_j +
A_{ij}\, q_ip_j - \Delta = 0\;, \label{quadratic1}\ee
where $S_{ij}$ is a symmetric matrix, $A_{ij}$ an anti-symmetric
matrix, $\Lambda$ a constant with dimensions $[L^2/(\textrm{Action})]^2$,
and $\Delta$ a real constant.  This class includes a number
of interesting cases.
For example, in geometrodynamics, the function $C_{\vec N}$
obtained by smearing the diffeomorphism constraint with any
vector field $\vec{N}$ on the `spatial' manifold $M$,
\be C_{\vec N}(q,p)\, = \,\int_{M} P^{ab}(x)\, {\pounds}_{\vec N}
\,q_{ab}(x)\, \dd^3x\;, \ee
is of this type. (In the `super-index' notation introduced by
Bryce DeWitt, $i \equiv (x; a,b)$, this function is of the type
$C_{\vec N}= q_i N_{ij} p_j$ where $N_{ij}$ is an anti-symmetric
matrix. Thus, in this example, $S_{ij}=0$ and $A_{ij} = N_{ij}$.)
As we will see in sections \ref{s5.3} and \ref{s5.4}, some of the
finite-dimensional examples that have been studied extensively in
the literature are also of this type. For simplicity, in the
detailed analysis we will consider a single constraint, but it is
rather straightforward to extend it to allow for a set of
commuting constraints each of which is of type (\ref{quadratic1}).

\subsection{Setup and physical states}
\label{s5.1}

\noindent In terms of the holomorphic coordinates $z_i$ of Eq
(\ref{zi}), the constraints we consider can be written as
\be C(q_i,p_i) := \kappa_{ij}z_i \bar{z}_j - \Delta =0\;,
\label{quadratic3}\ee
where $\kappa_{ij}$ is a Hermitian matrix (and we have chosen all
$\ell_i$ equal to one another in the definition \ref{zi} of
holomorphic functions $z_i$).
As in section \ref{s4.1}, the analysis becomes more
transparent if the initial choice of canonical coordinates is
adapted to the constraint at hand. Let us therefore choose
$(q_i,p_i)$ such that the Hermitian matrix $\kappa_{ij}$ is
diagonal, with eigenvalues $\kappa_i$ (which are of course real).
Then, using as before normal ordering, the quantum constraint
operator becomes
\be \h{C} = \kappa_j\,\h{N}_j - \Delta\,\h1\;, \label{quadratic2}\ee
with $\h{N}_j$ the $j$th number operator, $\h{N}_j =
\h{a}_j^\dag \h{a}_j$ (where there is no summation over $j$).

The action of $\h{U}(\lambda) := \rme^{-\ii\lambda \h{C}}$ on the
kinematical coherent states $\Psi_\a$ can be calculated in a
straightforward manner:
\ba \rme^{-\ii\lambda\h{C}}\,\ket{\Psi_\a} &=&
\rme^{\ii\lambda\Delta}\,
\rme^{-\ii\lambda\Sigma_j\kappa_j\h{a}_j^\dag\h{a}_j}\,\bigotimes_{i=1}^D
\bigg[\rme^{-|\alpha_i|^2/2} \sum_{n_i=0}^\infty
\frac{(\alpha_i)^{n_i}}{\sqrt{n_i!}} \,\ket{n_i}\bigg] \nonumber \\
&=& \rme^{\ii\lambda\Delta} \bigotimes_{i=1}^D \bigg[
\rme^{-|\alpha_i|^2/2}\,
\sum_{n_i=0}^\infty \frac{(\rme^{-\ii\lambda\kappa_i}
\alpha_i)^{n_i}}{\sqrt{n_i!}} \ket{n_i}\bigg]\nonumber\\
&=& \rme^{\ii\lambda\Delta}\, \ket{\Psi_{\a(\lambda)}}\;,
\vphantom{\bigotimes^D} \label{expc} \ea
with $\a_j(\lambda) = \rme^{-\ii\lambda\kappa_j}\a_j$ (where there
is no summation over $j$). Thus, apart from a constant phase
factor, the image of $\Psi_\a$ under $\h U(\lambda)$ is again a
coherent state, the peak being shifted from $\a$ to $\a(\lambda)$.
It is easy to verify that, on the classical phase space $\G$,
$\a(\lambda)$ is precisely the gauge orbit passing through $\a$,
generated by the constraint function $C$. Consequently, as in
section \ref{s4}, there is a close interplay between classical and
quantum theories also for the quadratic constraints now under
consideration: the action of $\h{U}(\lambda)$ simply moves the
peak of the kinematical coherent state along the gauge orbit
generated by $C$ on $\G$.

Since we wish to use the group averaging technique, we need to
restrict ourselves to the case in which $\h{U}(\lambda)$ provides
a representation of a group on $\H\kin$. The expression
(\ref{expc}) for the action of $\h{U}(\lambda)$ shows that it
provides a representation of U(1) if and only if there is a real
number $\Lambda$ such that $\rme^{\ii \Lambda \Delta} = \rme^{-\ii
\Lambda \kappa_i} = 1$ for all $i$. We will consider constraints
of this type. This in turn guarantees that the kernel of $\h{C}$,
the physical Hilbert space $\H\phy$, is a subspace of $\H\kin$,
and group averaging now falls in the simpler of the two cases
considered in section \ref{s2.2}, defining a projection operator
$P$ from $\H\kin$ to $\H\phy$. The condition also means that all
ratios $\kappa_i/\Lambda$ and $\kappa_i/\kappa_j$ have rational
values, and since we can multiply $C$ by a constant, we may take
all $\kappa_i$ as well as $\Delta$ to be integers, and $\Lambda =
2\pi$ (choosing the integers with the smallest absolute values
fixes them uniquely up to an overall sign). In addition, it can be
seen from the form (\ref{quadratic2}) of $\h{C}$ that, in order
for $\h C$ to have a non-trivial kernel, $\kappa_i$ and $\Delta$
must be such that $\kappa_i n_i -\Delta =0$ for \emph{some} choice
of integers $n_1,\ldots n_D$. To ensure that the example is
interesting, we will assume this to be the case.

The physical state corresponding to $\Psi_\a$ is then given by
Eq (\ref{proj}), or
\be \Psi_\a^{\textrm{phy}} = \frac{1}{2\pi}\, \int_0^{2\pi}
\dd\lambda\, \h{U}(\lambda) \Psi_\a\;, \ee
and using (\ref{ip1}) and (\ref{sprod}) we find that its norm is
\ba \Vert\Psi_\a^{\textrm{phy}}\Vert^2 &=& \frac{1}{2\pi}\,
\int_0^{2\pi}\dd\lambda\, \ev{\h{U}(\lambda)\Psi_\a|\Psi_\a}\nonumber\\
&=& \frac{\rme^{-|\a|^2}}{2\pi}\, \int_0^{2\pi} \dd\lambda\,
\rme^{-\ii\lambda\Delta}\,\, \rme^{\Sigma_j
|\a_j|^2\,\rme^{\ii\lambda\kappa_j}}\,. \label{norm2}\ea
Since the integrand is a smooth function of $\lambda$ and the
integral is over a closed interval, it is clearly well-defined. By
expanding the coherent states in the first equation of
(\ref{norm2}) in the occupation number basis, the norm can also be
expressed as the sum
\be  \Vert\Psi_\a^{\textrm{phy}}\Vert^2 = \rme^{-|\a|^2}
\sum_{n_1,\ldots, n_D}\, \frac{|\a_1|^{2n_1}\ldots |\a_D|^{2n_D}}
{n_1!\ldots n_D!}\,\delta_{\kappa_i n_i,\, \Delta}\;,
\label{norm3} \ee
where the Kronecker delta limits the contribution to a finite set
of $n_i$, those satisfying $\kappa_i n_i = \Delta$. In this series
form, positivity of the norm is manifest, whence the result of
group averaging is indeed a physical state. More generally, as
stated in section \ref{s2.2}, one can verify that the space $\S$
spanned by finite linear combinations of coherent states is an
admissible dense sub-space of $\H\kin$ to serve as the `seed' in
the group averaging procedure.

Notice that, if the $\kappa_i$ are all non-negative, one can also
obtain a closed form for the norm directly from the last integral
in (\ref{norm2}). Setting $\z = \rme^{\ii\lambda}$, the integral
reduces to one over the unit circle in the complex $\z$ plane,
which can be evaluated using the method of residues:
\be \Vert\Psi_\a^{\textrm{phy}}\Vert^2
= \frac{\rme^{-|\a|^2}}{2\pi \ii}\,
\oint_{|\z| =1} \dd\z \,\frac{\rme^{\Sigma_j |\a_j|^2
\z^{\kappa_j}}}{\z^{\Delta +1}}
= \frac{\rme^{-|\a|^2}}{\Delta\, !}\,
\frac{\dd^\Delta}{\dd\z^{\Delta}} \,\rme^{\Sigma_j |\a_j|^2
\z^{\kappa_j}}\bigg|_{\z=0}\,, \ee
from which the result (\ref{norm3}) can be recovered by evaluating
the derivative. If all the $\kappa_i$ are non-positive, one can
use the same procedure by initially setting $\z= \rme^{-\ii
\lambda}$. In the general case, however, this method gives a
result expressed only as an infinite series.

\subsection{Physical observables}
\label{s5.2}

\noindent The calculations of expectation values and fluctuations
of Dirac observables can also be carried out in a rather
straightforward manner, because the expressions involve integrals
of the type $\int \dd\lambda\,
\ev{\Psi_{\a(\lambda)}|\h\cO|\Psi_\a}$ in which both the bra and
the ket are coherent states. However, in the general case the
result can again only be expressed as a well-defined integral or a
convergent infinite series. Properties of expectation values and
fluctuations, therefore, are not always transparent. But in the
case of linear or quadratic observables, some interesting results
can be readily obtained.

A linear observable is one of the form $\cO = \bar{F}_iz_i + F_i
\bar{z}_i$, and calculations of expectation values and
fluctuations with it are greatly simplified by the requirement
that it be a strong Dirac observable. From the Poisson bracket
between $\cO$ and the constraint (\ref{quadratic3}) we see that in
this case $F_i = 0$ for all $i$ such that $\kappa_i \not= 0$.  In
a group averaged coherent state,
\ba \bra{\Psi_\a^{\textrm{phy}}}\,\h\cO\,\ket{\Psi_\a^{\textrm{phy}}}
&=& \frac{1}{2\pi} \int_0^{2\pi}\dd\lambda\,\rme^{-\ii\lambda
\Delta}\,\bra{\Psi_{\a(\lambda)}}\,F_i\ha_i^\dag + \bar F_i\ha_i\,
\ket{\Psi_\a} \nonumber\\
&=& \frac{1}{2\pi} \int_0^{2\pi}\dd\lambda\,\rme^{-\ii\lambda
\Delta}\,[F_i\,\bar\a_i(\lambda) + \bar F_i\,\a_i]\,
\ev{\Psi_{\a(\lambda)}|\Psi_\a}\;. \label{obs1}\ea
But $F_i\,\bar\a_i(\lambda) =
F_i\,\rme^{\ii\kappa_i\lambda}\bar\a_i = F_i\,\bar\a_i$, since for
each $i$ either $F_i$ or $\kappa_i$ vanish (or both). Therefore,
\be \ev{\h\cO}_\a^{\textrm{phy}}
= \frac{\bra{\Psi_\a^{\textrm{phy}}}\,\h\cO\,
\ket{\Psi_\a^{\textrm{phy}}}}{\Vert
\Psi_\a^{\textrm{phy}}\Vert^2} = \cO(\a)\;, \ee
the classical value. Similarly, for the fluctuation we calculate
\ba \bra{\Psi_\a^{\textrm{phy}}}\,\h\cO^2\ket{\Psi_\a^{\textrm{phy}}}
&=& \frac{1}{2\pi} \int_0^{2\pi}\dd\lambda\,\rme^{-\ii\lambda
\Delta}\,\bra{\Psi_{\a(\lambda)}}\,(F_i\ha_i^\dag + \bar
F_i\ha_i)^2
\,\ket{\Psi_\a} \nonumber\\
&=& \frac{1}{2\pi} \int_0^{2\pi}\dd\lambda\,\rme^{-\ii\lambda
\Delta}\,[F_iF_j\,\bar\a_i(\lambda)\bar\a_j(\lambda) + 2\,F_i\bar
F_j\, \bar\a_i(\lambda)\a_j + \nonumber\\ & &\qquad\qquad\qquad +\
\bar F_i\bar F_j\,\a_i\a_j + F_i\bar F_i]\,
\ev{\Psi_{\a(\lambda)}|\Psi_\a} \nonumber\\ \vphantom{\int^A} &=&
[(F_i\bar\a_i + \bar F_i\a_i)^2 + F_i\bar F_i]\,
\Vert\Psi_\a^{\textrm{phy}}\Vert^2\;, \ea
where the last term in the second expression arises from the
commutator $[\ha_i,\ha_j^\dag]$, and $F_i\,\bar\a_i(\lambda) = F_i
\,\bar\a_i$ for each $i$, for the same reason as in the
expectation value. From this, it follows immediately that
\be (\Delta\h\cO)\phy^2 = (\Delta\h\cO)\kin^2 = F_i\bar F_i\;.\ee

In the case of a quadratic observable of the general form $\cO =
\phi_{ij}\,z_i\bar z_i$, one can use similar reasoning to arrive
at a more limited statement. Since the Poisson bracket of $\cO$
with the constraint (\ref{quadratic3}) vanishes, the matrices
$\kappa_{ij}$ and $\phi_{ij}$ commute. Therefore, they can be
simultaneously diagonalized and we can write the normal ordered
operator for the quadratic observable as $\h\cO = \phi_i\,\h N_i$.
Then
\ba \bra{\Psi_\a^{\textrm{phy}}}\,\h\cO\,
\ket{\Psi_\a^{\textrm{phy}}} &=& \frac{1}{2\pi} \sum\nolimits_i
\int_0^{2\pi}\dd\lambda\,\rme^{-\ii\lambda
\Delta}\,\bra{\Psi_{\a(\lambda)}}\,\phi_i\,\ha_i^\dag\ha_i\,
\ket{\Psi_\a} \nonumber\\
&=& \frac{1}{2\pi} \sum\nolimits_i
\int_0^{2\pi}\dd\lambda\,\rme^{-\ii\lambda
\Delta}\,\phi_i\,\rme^{\ii\lambda\kappa_i}\bar\a_i\a_i\,
\ev{\Psi_{\a(\lambda)}|\Psi_\a}\;. \ea
In this expression, the $\lambda$ dependence of the integrand
cannot be simplified as in the corresponding one for a linear
observable, (\ref{obs1}), and the expectation value
$\ev{\h\cO}_\a^{\textrm{phy}}$ is typically different from $\cO(\a)$.
We will analyze the difference in detail in two examples. Here, we
will only present a general argument to show that, when the phase
space point $\a$ at which $\Psi_\a$ is peaked is far from the
origin, the expectation values
\be \ev{\h\cO}_\a^{\textrm{phy}}
= \frac{\bra{\Psi_\a^{\textrm{phy}}}\,\h\cO\,
\ket{\Psi_\a^{\textrm{phy}}}} {\Vert
\Psi_\a^{\textrm{phy}}\Vert^2} \label{expval} \ee
are necessarily close their classical values. For simplicity, let
us explicitly consider the case in which one of the $|\a_i|^2$
becomes very large; without loss of generality, we may assume it
is $|\a_D|^2$.

Recalling that $\ev{\Psi_{\a(\lambda)} | \Psi_\a} = \rme^{-|\a|^2}
\, \rme^{\Sigma_j|\a_i|^2\exp(\ii \lambda\kappa_j)}$, we can write
the numerator of (\ref{expval}) as
\be \bra{\Psi_\a^{\textrm{phy}}}\,\h\cO\,\ket{\Psi_\a^{\textrm{phy}}}
= \frac{\rme^{-|\a|^2}}{2\pi} \sum\nolimits_i \phi_i|\a_i|^2
\int_0^{2\pi}\dd\lambda\,\,\rme^{-\ii\lambda(\Delta-\kappa_i)}\,
\rme^{\Sigma_j|\a_j|^2\,\rme^{\ii\lambda\kappa_j}}. \ee
Each integral in this last expression is of the form (\ref{norm2})
with $\Delta$ replaced by $\Delta-\kappa_i$, and can therefore be
written immediately in the form of the result (\ref{norm3}); thus,
\be \bra{\Psi_\a^{\textrm{phy}}}\,\h\cO\,\ket{\Psi_\a^{\textrm{phy}}}
= \rme^{-|\a|^2} \sum_{i=1}^D \left(\phi_i|\a_i|^2 \sum_{n_1,\ldots,
n_D}\, \frac{|\a_1|^{2n_1}\ldots |\a_D|^{2n_D}} {n_1!\ldots
n_D!}\,\delta_{\kappa\cdot n,\, \Delta-\kappa_i} \right).
\label{obs3} \ee
Both the numerator (\ref{obs3}) and the denominator (\ref{norm3})
of the expectation value are finite polynomials in the $|\a_i|^2$,
which implies that when $|\a_D|^2 \to \infty$ the dominant terms
in the summations over $n_1,...,n_D$ are those with the highest
powers of $|\a_D|^2$. Thus, to determine the asymptotic value of
$\ev{\h\cO}_\a^{\textrm{phy}}$ we need to find the highest values of
$n_D$ contributing to the sums in (\ref{obs3}) and (\ref{norm3}).

Call $N_D$ the highest value of $n_D$ contributing to the
norm (\ref{norm3}). The classical constraint implies that, as
$|\a_D|^2 \to \infty$, $\Delta \sim \kappa_D|\a_D|^2$; together
with the restriction on the $n_i$, this then implies that $N_D
\sim |\a_D|^2$.  Now turn to (\ref{obs3}) and consider first a
term with $i\ne D$. Depending on the values of all the $\kappa_j$,
the highest value of $n_D$ in it may be either $N_D$ or a smaller
number; in the former case, the whole term is of order 1 in
$|\a_D|^2$, in the latter it has a decreasing behavior. For
the $i=D$ term in (\ref{obs3}) the situation is simpler, since
the highest value of $n_D$ in it is always $N_D - 1$; in fact,
the leading term in the sum is $|\a_D|^{2(N_D-1)}/(N_D-1)!$
times the same combination of the remaining $\a_j$ and $n_j$
that multiplies $|\a_D|^{2N_D}/N_D!$ in the norm (\ref{norm3}).
Thus, as $|\a_D|^2 \to \infty$,
\be \ev{\h\cO}_\a^{\textrm{phy}} = \frac{\bra{\Psi_\a^{\textrm{phy}}}
\, \h\cO\,\ket{\Psi_\a^{\textrm{phy}}}}{\Vert\Psi_\a^{\textrm{phy}}
\Vert^2} \sim \phi_D\,|\a_D|^2\, \frac{|\a_D|^{2(N_D-1)}/(N_D-1)!}{
|\a_D|^{2N_D}/N_D!} \sim \phi_D\,|\a_D|^2 \sim \cO(\a)\;. \ee

Two examples discussed in detail below will illustrate this
general feature of our physical states. For some quadratic
observables $\ev{\h\cO}_\a^{\textrm{phy}}$ will actually coincide with
the classical value $\cO(\a)$. These examples will also show that
the fluctuations of quadratic observables in the physical states
differ from the corresponding fluctuations in the kinematical
states even for the simplest observables, but do so in a
controlled fashion. Interestingly, in some cases they are smaller
than the latter.

\subsection{Example 1: Constrained total energy}
\label{s5.3}

\noindent In our first example, we will consider two coupled
harmonic oscillators of the same mass and spring constant, subject
to the constraint that the sum of their energies have a fixed
value:
\be \tilde{C} := \sum\nolimits_i \frac{p_i^2}{2m} + k q_i^2 -
\tilde{\Delta} = 0\;. \ee
This example has drawn attention in the literature (see, e.g.,
Refs \cite{Rovelli,at,Ashworth}) because a constraint that fixes
the total energy provides a toy model for the Hamiltonian
constraint of general relativity in the spatially compact case
(where the energy is zero), and can be used to study the
associated `problem of time'.

To conform to our general framework, we need to divide $\tilde{C}$
by a constant with dimensions of energy to obtain a dimensionless
constraint function. While any constant will do, we will use the
natural choice $\hbar \omega$. Then, using for $\ell$ in the
expression (\ref{zi}) of $z_i$ the natural length scale $\ell =
\sqrt{\hbar/m\omega}$, the constraint can be re-expressed as
\be C := \frac{1}{\hbar \omega}\; \tilde{C}
= z_1\bar{z}_1+z_2\bar{z}_2- \Delta = 0\;, \ee
where $\Delta = \tilde{\Delta}/\hbar\omega$. The kinematic phase
space $\G$ is $\mathbb{R}^4$; the constraint surface $\bar\G$ is a
3-sphere; and the gauge orbits generated by the constraint
function ${C}$ provide a Hopf fibration of $\bar\G$. Thus, the
reduced phase space $\h{\G}$ is a 2-sphere. Because of the
topological non-triviality, although $\h{\G}$ is 2-dimensional, we
need a set of at least three Dirac observables to separate points
of $\hat\G$. A convenient choice is
 \be L_1 = \textrm{Re}\,\, z_1\bar{z_2},\quad
 L_2 = \textrm{Im}\,\, z_1\bar{z_2}, \quad
 L_3 = \half\,(z_1\bar{z}_1 - z_2 \bar{z}_2)\;. \ee
(As the notation suggests, they are also closed under Poisson
brackets, providing a representation of the standard angular
momentum algebra. However, this property will not play a role in
our semi-classical considerations.)

In contrast to section \ref{s4}, now the structure of Dirac
observables is quite complicated and, as seen in the general
discussion, there is no a priori reason to suppose that expectation
values and fluctuations of these observables in the kinematic and
physical quantum states will be simply related. To explore the
relation, let us begin by constructing physical states. Our
constraint operator is
\be \h{C} = \ha_i\ha_i^\dag - \Delta
= \sum\nolimits_i \h{N}_i - \Delta\;. \ee
Thus, $\kappa_i = 1$ for $i=1,2$. We can now write down the
physical coherent states (\ref{expc}):
    \be
    \ket{\Psi^{\textrm{phy}}_\a}
    = \frac{\rme^{-(|\alpha_1|^2+|\alpha_2|^2)/2}}{2\pi}
    \int_0^{2\pi} \dd\lambda\,
    \sum_{n,m=0}^\infty
    \frac{\alpha_1^{n}\,\alpha_2^m}{\sqrt{n!}\, \sqrt{m!}}\,
    \rme^{-\ii\lambda\hat C}\,
    \ket{n,m}\;.
\ee
Using the fact that the Fock basis $\ket{n,m}$ is an
eigenbasis for the constraint, satisfying $\hat C \ket{n,m} =
n+m-\Delta$, where $\Delta=k:= n+m$ is an integer, we get
\be
    \ket{\Psi^{\textrm{phy}}_\a} = \frac{1}{2\pi}\,\int_0^{2\pi}
    \dd\lambda\, \rme^{\ii\lambda\Delta}\;
    \ket{(\alpha_1\rme^{-\ii\lambda}),(\alpha_2\rme^{-\ii\lambda})}\;.
\ee

Hence, from the general discussion of section \ref{s5.1}, we
conclude that the norm of the physical states $\Psi_\a^{\textrm{phy}}$,
obtained by group averaging the coherent states $\Psi_\a$, is
given by
\ba
   \Vert\Psi^{\textrm{phy}}_\a\Vert^2
   &=& \frac{\rme^{-|\a|^2}}{2\pi \ii}\oint_{|\zeta|=1}
   \dd\z \;\frac{\rme^{|\a|^2\z}}{\z^{\Delta+1}}\nonumber\\
   &=& \frac{|\a|^{\Delta} \rme^{-|\a|^2}}{\Delta\,!}\;.
\ea
If we now use the fact that the coherent state is chosen to be
peaked at a point of the constrained surface $\bar{\G}$, then $\Delta
= |\alpha|^2 = k$, so we have $\Vert\Psi^{\textrm{phy}}_\a\Vert^2 =
\rme^{-k}\,k^k/k!$ . Note that in this case, the physical Hilbert
space is finite-dimensional (due to the compactness of the reduced
phase space), with dim$({\cal H}\phy) = k+1$, and that the
extractor operator $\hat E = \hat P$ is in fact a projection
operator on the kinematical Hilbert space, whose action is to
project the kinematical coherent state to the subspace spanned by
kets of the form $\ket{n,k-n}$ for a fixed value of $k$.

We can now calculate the expectation values of the constraint and
the observables $L_I$, $I=1$, 2, 3.%
\footnote{The expectation value of the constraint in the
kinematical coherent state vanishes, and its fluctuation is given
by
\be (\Delta\hat C)\kin^2
= |\alpha|^2=|\alpha_1|^2+|\alpha_2|^2=\Delta\;.\ee
Since $(\Delta\hat C)\kin^2$ is proportional to $E\cl/\hbar \w$,
for coherent states with a large value of $E\cl$, the fluctuation
of the constraint will also be large.}
Let us begin with the term $a_1^\dag a_1$, and compute its
expectation value $\bra{\Psi_\alpha}\,a_1^\dag
a_1\ket{\Psi_\alpha}\phy$. In this case we have
\ba
    \langle\Psi_\alpha|\,a_1^\dag a_1\ket{\Psi_\alpha}\phy &=&
    \frac{1}{2\pi}\,\int_0^{2\pi} \dd\lambda\,
    \rme^{-\ii\lambda\Delta}\;\langle(\alpha_1\rme^{-\ii\lambda}),
    (\alpha_2\rme^{-\ii\lambda})|\,a_1^\dag\,a_1\,
    \ket{\alpha_1,\alpha_2}\nonumber\\
 &=&\frac{|\alpha_1|^2}{2\pi}\,\int_0^{2\pi} \dd\lambda\,
    \rme^{-\ii\lambda(\Delta-1)}\,\rme^{-|\alpha|^2}\,
    \rme^{\rme^{\ii\lambda}|\alpha|^2}\, ,
\ea
which can be rewritten as
\be
    \langle\Psi_\alpha|\,a_1^\dag a_1\ket{\Psi_\alpha}\phy
    =|\a_1|^2\,
    \frac{\rme^{-\tilde E}}{2\pi\ii}\oint_{|\z|=1}
    \frac{\dd\z}{\z}\,\z^{-(\tilde E-1)}\, \rme^{\tilde E\,\z}
    = |\a_1|^2\,\frac{\rme^{-k}}{(k-1)!}\, k^{k-1}\, .
\ee
Thus, when we compute the expectation value we get
\be
     \frac{\langle\Psi_\alpha|\,a_1^\dag a_1\ket{\Psi_\alpha}\phy}{
     \langle\Psi_\alpha\ket{\Psi_\alpha}\phy}
     =|\alpha_1|^2\,\frac{\rme^{-k}}{(k-1)!} \, k^{k-1}\;
     \frac{k!}{\rme^{-k}}\; k^{-k} = |\alpha_1|^2\,.
\ee
It is clear that the expectation value of the operator
$a_2^\dag\,a_2$ will also equal its classical value. Finally, it
is easy to show that the expectation value of the observables
$L_I$ in the physical coherent state labelled by
$(\alpha_1,\alpha_2)$ is given, for all $I$, by
\be
     \langle \hat L_I\rangle\phy:=
     \frac{\langle\Psi_\alpha|\,\hat L_I\ket{\Psi_\alpha}\phy}
      {\langle\Psi_\alpha\ket{\Psi_\alpha}\phy}
     = L_I\big|\cl \, \frac{k}{|\alpha_1|^2+|\alpha_2|^2}\;.
\ee
However, since the point $\alpha$ lies on the constraint surface,
we have $|\alpha_1|^2+|\alpha_2|^2 = k$, whence the expectation
values always coincide with the classical values.

The next step is to look at the fluctuations of Dirac
observables. We shall first compute them in the physical coherent
states, and later compare them to the fluctuations in the
kinematical coherent states. It is a straightforward calculation
to show that, on a physical coherent state, \be
     \ev{\hat L^2_I}\phy = L^2_I\big|\cl\,
     \frac{k(k-1)}{(|\alpha_1|^2+|\alpha_2|^2)^2}+ \frac{k}{4}\;.
\ee
Therefore,
\be
    (\Delta\hat L_I)\phy^2
    := \ev{\hat L^2_I}\phy - (\ev{\hat L_I}\phy)^2
    = L^2_I\big|\cl
    \left(\frac{-k}{(|\alpha_1|^2+|\alpha_2|^2)^2}\right)
    + \frac{k}{4} =
    -\frac{1}{k}\;L^2_I|\cl
    + \frac{k}{4} \;.
\ee
On the other hand, the fluctuations in the kinematical
coherent states are given by
\be
(\Delta\hat L_I)\kin^2
= \frac{1}{4}\,(|\alpha_1|^2+|\alpha_2|^2)=\frac{\Delta}{4}\; .
\ee
Thus, the difference between the fluctuations is given by
\be
    (\Delta\hat L_I)\kin^2- (\Delta\hat L_I)\phy^2=\frac{1}{k}
    \;L^2_I\big|\cl < (\Delta\hat C)\kin^2\;.
\ee
This implies that: i) The difference in the fluctuations is
smaller than the fluctuation of the constraint operator on
$\H\kin$; ii) Group averaging actually reduces the
dispersions. To summarize, then, if we begin with semi-classical
kinematic states peaked at points on the constraint surface,
physical states resulting from group averaging are
\emph{guaranteed} to be semi-classical. Furthermore, the
kinematical calculation provides a good upper bound on the
dispersion in the physical states.

Let us conclude with two remarks. \\
1. While the explicit calculations led to a desirable result,
the `mechanism' behind its success is rather obscure, even in
retrospect. Indeed, from the expressions of the individual
integrals or summations defining the numerators $\ev{
\Psi^{\textrm{phy}}_\a\,|\,\hat{L}_i\,|\,\Psi^{\textrm{phy}}_\a}$
and the denominators $\ev{\Psi^{\textrm{phy}}_\alpha|
\Psi^{\textrm{phy}}_\alpha}$ in the expression of the expectation
values, it is far from clear that the ratio would agree so well with
the result on the kinematical Hilbert space. It is only after the two
integrals are evaluated explicitly that the required cancellations
occur. Thus, the desired result emerges somewhat surprisingly and
only in the very final step.\\
2. We could have chosen the quantum constraint $\hat C$ not to be
normal ordered. Had we used the standard text-book ordering, even
in the kinematical Hilbert space the expectation value of the
constraint would not have been zero if the coherent states were
chosen to be peaked at points $\a$ on $\bar\G$: Due to the zero
point energy, we would then have $\ev{\h C}\kin =  1$. Furthermore,
the expectation values of the Dirac observables on group averaged
coherent states would not have coincided with the classical
values, but would have the form $\ev{\hat L_i}\phy = L_i|\cl\,
(1-\hbar\w/E\cl).$ Thus, only for large values of ${E\cl}$
would the physical coherent states have approximated the
classical values of the observables. These `discrepancies' can be
avoided by choosing kinematical coherent states which are peaked
slightly away from the constraint surface, but that would have
made the construction a bit obscure from a physical standpoint.
Fluctuations would have slightly different values but share the
same qualitative behavior as in the normal ordered case.

\subsection{Example 2: Constrained energy difference}
\label{s5.4}

\noindent In our second example, we will consider two coupled
harmonic oscillators of the same mass and spring constant, subject
to the constraint that the difference of their energies have a
fixed value:
\be \tilde{C} :=  \left(\frac{p_1^2}{2m} + k q_1^2\right)
-\left(\frac{p_2^2}{2m} + k q_2^2\right) - \tilde{\Delta}=0\;.\ee
This example has also drawn attention in the literature (see,
e.g., Refs \cite{at,Ashworth}) because it models some cosmological
situations \cite{at}. The system is conceptually simpler, even
though it is more complicated from the technical viewpoint, since
now the reduced phase space is non-compact and isomorphic to an
open set in $\mathbb{R}^2$. We again have two harmonic oscillators
with the same frequency, but now the constraint can be written as
\be
    C=|z_1|^2-|z_2|^2-\Delta = 0\;,
\ee
which is of the form (\ref{quadratic3}) with $\kappa_{ij} =
\textrm{diag}(1,-1)$, where as before $\Delta = \tilde\Delta/
\hbar\omega$. The quantum constraint operator has the form
\be
    \hat C = (\ha_1^\dag \ha_1 - \ha_2^\dag \ha_2)
    - \tilde{\Delta}\;.
\ee
(Note that, in this example, the ordering ambiguity discussed at
the end of the last sub-section does not matter because the
zero-point energies of the oscillators cancel each other out.) It
is straightforward to write the physical coherent states,
\be \ket{\Psi_{\alpha}^{\textrm{phy}}}
    = \frac{1}{2\pi}\,\rme^{-(|\alpha_1|^2+|\alpha_2|^2)/2}
    \int_0^{2\pi} \dd\lambda\,
    \sum_{n,m=0}^\infty
    \frac{\alpha_1^{n}\,\alpha_2^m}{\sqrt{n!}\, \sqrt{m!}}\,
    \rme^{-\ii\lambda\hat C}\,
    \ket{n,m}\;.
\ee
Using the fact that the Fock basis $\ket{n,m}$ is an eigenbasis
for the constraint operator, satisfying $\hat C\ket{n,m} = [(n-m)
- \Delta] \ket{n,m}$, where $\Delta=k:= n-m$ is an integer, we get
\be
   \ket{\Psi_{\alpha}^{\textrm{phy}}}
   = \frac{1}{2\pi}\,\int_0^{2\pi} \dd\lambda\,
   \rme^{\ii\lambda\Delta}\;\rme^{-(|\alpha_1|^2+|\alpha_2|^2)/2}
   \sum_{n=0}^{\infty}\frac{(\rme^{-\ii n\lambda}\alpha_1^n)\,(
   \rme^{\ii m\lambda} \alpha_2^{m})}
   {\sqrt{n!\, m!}}\, \ket{n,m}\;,
\ee
but since the last term in the integral is precisely a
coherent state, the physical coherent state can be written as
\be
   \ket{\Psi_{\alpha}^{\textrm{phy}}}
   = \frac{1}{2\pi}\,\int_0^{2\pi} \dd\lambda\,
   \rme^{\ii\lambda\Delta}\;\ket{(\alpha_1\rme^{-\ii\lambda}),
   (\alpha_2\rme^{\ii\lambda})}\;.
\ee
Although the physical Hilbert space is now infinite-dimensional,
since zero is a discrete point in the spectrum of $\hat{C}$, the
extractor $\hat E$ is again a projection operator; $\hat{E} = \hat
P$ on the kinematical Hilbert space $\H\kin$. Its action is
to restrict contribution only to kets of the form $\ket{m,k+m}$,
for a fixed value of $k$. It is now straightforward to compute the
norm of the physical coherent states,
\ba
    \Vert\Psi^{\textrm{phy}}_\a\Vert^2:=( \Psi_\alpha^{\textrm{phy}}
    |\Psi_\alpha\rangle &=& \frac{1}{2\pi}\,\int_0^{2\pi}
     \dd\lambda\,
    \rme^{-\ii\lambda\Delta}\;\langle\Psi_{\alpha(\lambda)}\,
    \ket{\Psi_{\alpha}}
    \nonumber\\
    &=& \frac{1}{2\pi}\,\int_0^{2\pi} \dd\lambda\,
    \rme^{-\ii\lambda
    \Delta}\,\rme^{-(|\alpha_1|^2+|\alpha_2|^2)}\,
    \rme^{(\rme^{\ii\lambda}\,|\alpha_1|^2+\rme^{-\ii\lambda}\,
    |\alpha_2|^2)}\,.
\ea
To evaluate the integral, we can again define a new variable
$\z = \rme^{\ii\lambda}$ and express the integral over $\lambda$
as a contour integral over the unit circle.
\be
   \Vert\Psi^{\textrm{phy}}_\a \Vert^2 = \frac{\rme^{-
   \Delta}}{2\pi\ii}\oint_{|\z|=1}\,\frac{\dd \z}{\z}\,\z^{-
   \Delta}\,\rme^{(|\alpha_1|^2\,\z+|\alpha_2|^2\,\z^{-1})}\,.
\ee
Since $\alpha$ is chosen to lie on the constraint surface, from
the expression of the eigenvalues of $\hat{C}$, it follows that
$\Delta$ is again an integer; $\Delta=k$. But now the function to
be integrated has a pole of infinite order at the origin, whence
we can not compute the integral as easily in the previous example.
But we can still express the result in terms of special functions.
Let us expand the integral as an infinite series
\be
     \Vert\Psi^{\textrm{phy}}_\a \Vert^2
     =  \rme^{-(|\alpha_1|^2+|\alpha_2|^2)}|\alpha_1|^{2k}
     \sum_{m=0}^{\infty}\left[ \frac{1}{k!\,(k+m)!}\right]|
     \alpha_1|^{2m}|\alpha_2|^{2m}
\ee
and use the identity
\be
     \sum_{n=0}^{\infty}\frac{(x/2)^{2n}}{n!\,(k+n)!}
     = \left(\frac{2}{x}\right)^k \textrm{I}_k(x) \;.
\ee
where $\textrm{I}_m$ is a modified Bessel function. Then, the norm
can be expressed as:
\be
    \Vert\Psi^{\textrm{phy}}_\a\Vert^2 =
    \rme^{-(|\alpha_1|^2+|\alpha_2|^2)}
    (|\alpha_1|/|\alpha_2|)^k\; \textrm{I}_k(2|\alpha_1| |\alpha_2|)\;,
\ee

As one might expect from the last example, we again have three
quadratic Dirac observables:
\be
     J_3:= \half\,(z_1\bar{z}_1+z_2\bar{z_2})\;, \qquad
     J_+:= z_1\,{z}_2\;, \qquad
     J_-:= \bar{z}_1\,\bar{z}_2 \;,
\ee
with their corresponding combinations, $J_1 = \frac{1}{2}
(J_++J_-)$ and $J_2 = \frac{\ii}{2}\,(J_+-J_-)$. These observables
provide a realization of the sl(2,$\mathbb{R}$) = su(1,1) Lie
algebra that has been much studied. While this structure is
completely analogous to that in the previous example, now there is
a key difference. The reduced phase space $\hat{\G}$ is no longer
compact but topologically $\mathbb{R}^2$. One can verify that
$J_1$ and $J_2$ suffice to separate its points. Therefore, we will
first focus just on these observables and, for completeness, also
discuss $J_3$ at the end.

\begin{figure}
  \includegraphics[angle=270,scale=.50]{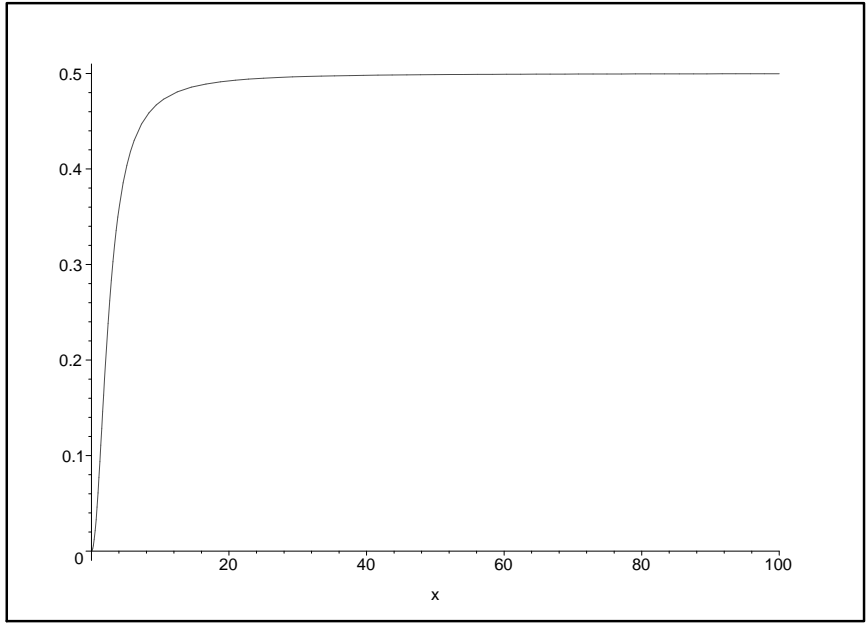}
\caption{\label{fig:1} The quotient $[(\Delta J_{1,2})^2\phy
-(\Delta J_{1,2})^2\kin]/(\Delta J_{1,2})^2\kin$, plotted as a
function of $x = |\a_2|$. The constant that fixes the energy
difference is set to $k=10$. Note that the ratio approaches 1/2,
and is very close to 1/2 even when $|\a_2|$ is not very large.}
\end{figure}

It is straightforward to verify that the expectation values of
$\hat J_\pm$ ---and therefore of $\hat J_{1,2}$---  in the
physical coherent states coincide with the classical values:
\be
     \ev{ \hat J_{1,2} }\phy = J_{1,2}|\cl \;.
\ee
For fluctuations we obtain
\be
     (\Delta\hat J_{1,2})^2\phy = \frac{|\a_1| |\a_2|}{4}
     \left[ \frac{\textrm{I}_{k-1}(2|\a_1||\a_2|) + \textrm{I}_{k+1}
     (2|\a_1| |\a_2|)} {\textrm{I}_{k}(2|\a_1| |\a_2|)} \right]
     + \frac{1}{2}
\ee
and
\be
     (\Delta\hat J_{1,2})^2\kin
     = \textstyle{\frac{1}{4}}\,(|\alpha_1|^2+|\a_2|^2+2)\;.
\ee
Therefore,
\be
     (\Delta\hat J_{1,2})^2\phy - (\Delta\hat J_{1,2})^2\kin
     = \frac{|\a_1| |\a_2|}{4}\,
     \left[\frac{\textrm{I}_{k-1}(2|\a_1||\a_2|)
     + \textrm{I}_{k+1}(2|\a_1||\a_2|)}{\textrm{I}_{k}(2|\alpha_1||\a_2|)}
     \right]-\frac{1}{4}\,(|\a_1|^2+|\a_2|^2)\;.
\ee

In Fig.~\ref{fig:1}, the fluctuations of $\h J_{1,2}$ in the
kinematical and physical states are compared. In particular,
the quotient
$$
   \frac{(\Delta J_{1,2})^2\phy - (\Delta J_{1,2})^2\kin}
   {(\Delta J_{1,2})^2\kin}
$$
is plotted as a function of $|\a_2|$ for $k =10$;
the qualitative behavior is the same for other values of $k$.
From the numerical investigations it is clear that the quotient
approaches $1/2$ very fast as $|\a_2|$ grows. This means that the
fluctuations in both types of coherent states are of the same
order. However, in this example, the fluctuations are smaller in
the kinematical coherent states than in the physical states and
the difference remains bounded away from zero. However, it is
clear that if the initial kinematical coherent states are chosen
with tolerances $2\delta_{i}/3$, $\epsilon_i$, we would be
guaranteed that the group averaged physical states will be
semi-classical with desired tolerances $\delta_i$ and
$\epsilon_i$.

\begin{figure}
  \includegraphics[angle=270,scale=.50]{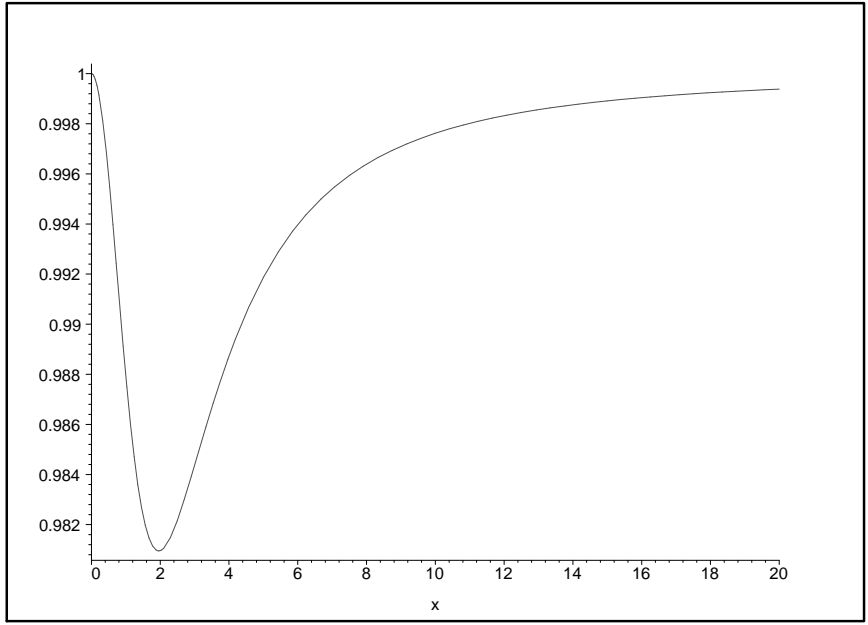}
  \caption{\label{fig:2} The quotient $\ev{\hat J_3}\phy/(J_3)\cl$,
plotted as a function of $x = |\a_2|$. The constant that fixes the
energy difference is set to $k=10$. Notice that the ratio is very
close to one even for values of $|\a_2|$ that are not that large.}
\end{figure}

Finally, for completeness, let us consider the quantum observable
$\hat J_3 = \frac{1}{2}\, (\ha_1^\dag\,\ha_1 + \hat a_2^\dag\,\hat
a_2)$. For the expectation value, we have:
\be
     \ev{\hat J_3}\phy = \frac{1}{2}\left[
     \,|\alpha_1|^2\left|\frac{\a_2}{\a_1}\right|
     \,\frac{\textrm{I}_{k-1}(2|\a_1| |\a_2|)}{|\a_1/\a_2|^k
     \,\textrm{I}_k(2|\a_1| |\a_2|)} +
     |\a_2|^2\left|\frac{\a_1}{\a_2}\right|
     \,\frac{ \textrm{I}_{k+1}(2|\a_1| |\a_2|)}{|\a_1/\a_2|^{k}
     \,\textrm{I}_k(2|\a_1| |\a_2|)}\right].
\ee
Since $|\alpha_1|^2=k+|\a_2|^2$, one can evaluate the last
expression as a function of $k$ and $|\a_2|$ and compare it with
the classical value. The quotient $(\ev{\hat
J_3}\phy)/(J_{3}|\cl)$ is plotted in Fig.~\ref{fig:2}. As can be
seen, already for small values of $k$ and $|\a_2|$, this quantity
is very close to one; physical coherent states approximate very
well the classical values of the Dirac observables we have chosen.

\begin{figure}
  \includegraphics[angle=270,scale=.50]{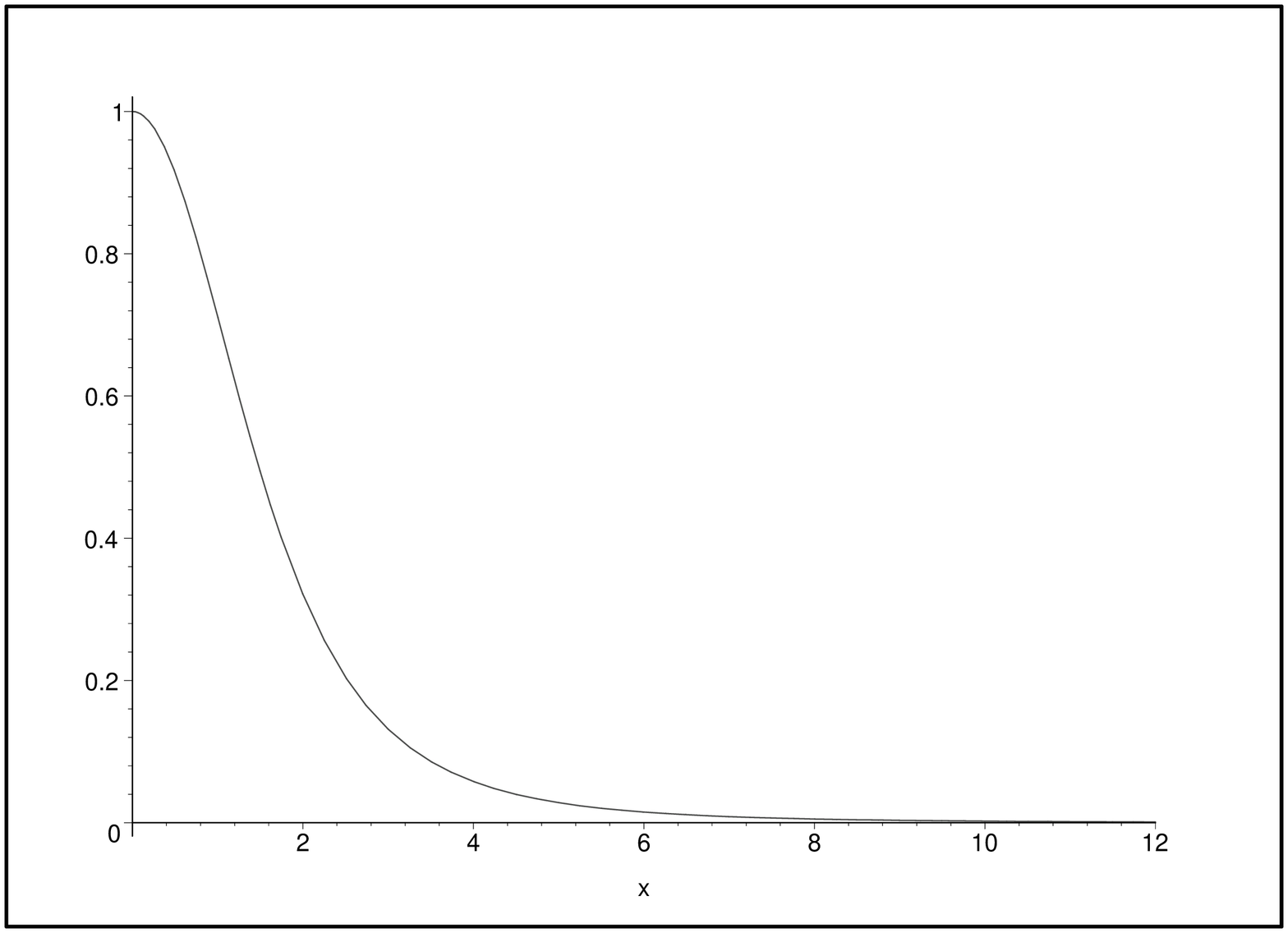}
  \caption{\label{fig:3} The quotient $[(\Delta\hat J_3)^2\kin-
(\Delta \hat J_3)^2\phy]/(\Delta \hat J_3)^2\kin$, plotted as a
function of $x=|\a_2|$. The constant that fixes the energy
difference is set to $k=10$. Note that the physical fluctuations
are smaller and very rapidly approach the kinematical ones.}
\end{figure}

The fluctuation of $\h J_3$ can be obtained by considering the
expectation value
\be
    \ev{\hat J^2_3}\phy
    = \frac{1}{4}\left[|\alpha_1||\a_2|\left(
    \frac{\textrm{I}_{k-1}
    +\textrm{I}_{k+1}}{\textrm{I}_{k}}\right)+|\alpha_1|^2|\a_2|^2\left(
    \frac{\textrm{I}_{k-2}+2\textrm{I}_{k}
    +\textrm{I}_{k+2}}{\textrm{I}_{k}} \right)\right].
\ee
With this expression we can again analyze the behavior of $(\Delta
\hat J_3)^2\phy$ and compare it with the kinematical fluctuation.
This comparison is shown in Fig.~\ref{fig:3}, where the quotient
$$
   \frac{(\Delta\hat J_3)^2\kin - (\Delta\hat J_3)^2\phy}
   {(\Delta\hat J_3)^2\kin}
$$
is plotted. Again the two fluctuations are of the same order. The
main difference with the fluctuations of $\hat{J}_1$, $\hat{J}_2$
is that now the fluctuations in the \emph{physical states} are
smaller but they very rapidly approach those in the kinematical
coherent states. Therefore the physical states are guaranteed to
be semi-classical with respect to $\hat{J}_3$. Thus, although the
detailed behavior of $\hat{J}_3$ is rather different from the one
of $\hat{J}_1$ and $\hat{J}_2$, the physical states are again
semi-classical also with respect to $\hat{J}_3$.

To summarize, by restricting the initial kinematical coherent
states to have suitably small tolerances, the group averaged
physical states can be guaranteed to be semi-classical for any
specified choice of tolerances.

\section{Discussion and Outlook}
\label{s6}

\noindent Let us begin with a brief summary of results. In section
\ref{s3} we clarified two issues concerning the notion of
semi-classical states. The clarifications in turn led us to a
criterion under which states $\Psi^{\textrm{phy}}_\a$ obtained by group
averaging suitable kinematical coherent states $\Psi_\a$ can be
regarded as semi-classical in $\H\phy$. In sections \ref{s4}
and \ref{s5} we saw that the criterion is satisfied if the
constraint sub-manifold $\bar\G$ is the level surface of a linear
function on $\Gamma$ or a quadratic function satisfying certain
conditions. Thus, the group averaging procedure offers a concrete
and potentially powerful strategy to construct \emph{physical}
semi-classical states for a class of constrained systems.

In the examples with quadratic constraints the result could not
have been foreseen on general grounds. Indeed, even in retrospect
we do not have a general understanding of `why' the strategy
works. It is particularly surprising that, in certain cases, the
group averaging procedure even \emph{reduces} the fluctuations of
Dirac observables. Now, all examples we considered have the
feature that $\hat{U}(\lambda) \Psi_\a$ is again a coherent state,
peaked at a point $\a(\lambda)$ of the phase space $\G$ on the
orbit of the gauge transformation generated by the classical
constraint $C$. This will not hold generally. Is this feature
perhaps the key to the `nicer than expected' behavior of the group
averaged coherent states? For example, since the value of any
classical Dirac observable $\cO$ is constant along a gauge orbit,
this feature implies that the expectation values $\ev{\hat{U}
(\lambda)\Psi_\a\,|\,\h\cO\,|\,\hat{U}(\lambda)\Psi_\a}$ are all
equal to the values $\cO(\a)$ of $\cO$ at the classical point $\a
\in \G$. Note, however, that it is not these expectation values
that dictate the calculation of $\ev{\Psi_\a^{\textrm{phy}}
\,|\,\h\cO\,|\,\Psi_\a^{\textrm{phy}}}$. As we saw in section \ref{s5},
the calculation is governed, rather, by a delicate interplay
between the \emph{cross} matrix elements $\ev{\Psi_\a \,|\,\h\cO
\,|\, \hat{U}(\lambda)\Psi_\a}$ and the norm of the physical state
$\Psi^{\textrm{phy}}_\a$. Neither of these by itself has any simple
relation to the value $\cO(\a)$ of the classical Dirac observable.
Thus, while the specific property now under consideration of our
class of constraints did simplify the detailed calculations,
it does not seem to suffice to ensure semi-classicality of
$\Psi^{\textrm{phy}}_\a$. Indeed, there are examples of quadratic
constraints ---such as those in the Bianchi I model--- which do not
share this property but where the group averaging procedure is again
useful in constructing physical semi-classical states \cite{bbc}. It
would be extremely useful to understand the underlying mechanism
which makes the group averaging strategy successful in a rather
diverse class of examples and prove a general result which
guarantees success for a wide class of constraints.

The following heuristics suggest that this may well be possible.
In the case of a generic constraint, one can simply choose the
constraint function itself as the first canonical coordinate, say
$q_1$, on the linear phase space $\G$, and then supplement it with
other suitably chosen functions (tailored to one's choice of Dirac
observables) to form a canonical coordinate system $(q_i,p_i)$.
Then, as in section \ref{s4}, physical states would have a simple
distributional form in the $q$-representation: $\Psi^{\textrm{phy}}
(q)=\delta(q_1) f(q_2,...,q_D)$. However, in the general case, this
canonical chart would not be adapted to the linearity of the
phase space, whence it would be difficult to identify  coherent
states in this representation. Nonetheless, it should be possible
to introduce \emph{some} notion of kinematical semi-classical
states $\Psi^{\textrm{kin}}_\a (q)$. Then the group averaging procedure
would lead to states $\Psi^{\textrm{phy}}_\a (q)= \delta(q_1)
\Psi^{\textrm{kin}}_\a (q_2,...,q_D)$. These are natural candidates for
physical semi-classical states. For, Dirac observables would act only on
the variables $q_2,\ldots, q_D$, whence there could be a simple
relation between their expectation values in $\Psi^{\textrm{kin}}_\a
(q)$ and $\Psi^{\textrm{phy}}_\a(q)$ needed to establish their
semi-classicality. A number of non-trivial technical problems need
to be overcome to determine whether these heuristic ideas can be
made precise. Nonetheless, they indicate that there may well be a
much more general underlying structure responsible for the success
of the group averaging method in the few examples discussed in
this paper.

\section*{ACKNOWLEDGEMENTS}

\noindent
This work was supported in part by NSF grants PHY-0010061,
PHY-0090091 and PHY-0354932, DGAPA-UNAM grant IN108103, CONACyT
grant 36581-E, the Alexander von Humboldt Foundation, the C.V.
Raman Chair of the Indian Academy of Sciences and the Eberly
research funds of Penn State.


\end{document}